\documentclass[aps, pre, a4paper, floatfix,  twocolumn, longbibliography, nofootinbib]{revtex4-1}

\usepackage[utf8]{inputenc}
\usepackage{epsfig}
\usepackage[T1]{fontenc}
\usepackage[english]{babel}
\usepackage[table]{xcolor}
\usepackage{t1enc}
\usepackage{graphicx}
\usepackage{amssymb}
\usepackage{amsmath}
\usepackage{relsize}
\usepackage[percent]{overpic}
\usepackage{bm}
\usepackage{hyperref}
\usepackage[section]{placeins}

\usepackage[normalem]{ulem}

\newcommand{\avg}[1]{{\left<#1\right>}}

\newcommand{\dd}{\mathrm{d}}

\usepackage{mathtools}
\def\multiset#1#2{\ensuremath{\left(\kern-.3em\left(\genfrac{}{}{0pt}{}{#1}{#2}\right)\kern-.3em\right)}}

\usepackage{amsmath}

\newcommand{\bb}{\bm{b}}
\newcommand{\A}{\bm{A}}
\newcommand{\D}{\bm{\mathcal{D}}}
\newcommand{\G}{\bm{G}}
\newcommand{\n}{\bm{n}}
\newcommand{\x}{\bm{x}}
\newcommand{\Q}{\bm{Q}}
\newcommand{\e}{\bm{e}}

\newcommand{\E}{\mathcal{E}}
\newcommand{\M}{\mathcal{M}}
\newcommand{\X}{\mathcal{X}}
\newcommand{\T}{\mathcal{T}}

\usepackage{verbatim}
\usepackage{overpic}
\usepackage{booktabs}
\usepackage{placeins}

\usepackage{siunitx}
\usepackage{multirow}

\usepackage{tikz}
\usetikzlibrary{external}
\begin{document}

\title{Reconstructing networks with unknown and heterogeneous errors}

\author{Tiago P. Peixoto}
\email{t.peixoto@bath.ac.uk}
\affiliation{Department of Mathematical Sciences and Centre for Networks
and Collective Behaviour, University of Bath, Claverton Down, Bath BA2
7AY, United Kingdom}
\affiliation{ISI Foundation, Via Chisola 5, 10126 Torino, Italy}

\begin{abstract}
The vast majority of network datasets contains errors and omissions,
although this is rarely incorporated in traditional network
analysis. Recently, an increasing effort has been made to fill this
methodological gap by developing network reconstruction approaches based
on Bayesian inference. These approaches, however, rely on assumptions of
uniform error rates and on direct estimations of the existence of each
edge via repeated measurements, something that is currently unavailable
for the majority of network data. Here we develop a Bayesian
reconstruction approach that lifts these limitations by not only
allowing for heterogeneous errors, but also for single edge measurements
without direct error estimates. Our approach works by coupling the
inference approach with structured generative network models, which
enable the correlations between edges to be used as reliable uncertainty
estimates. Although our approach is general, we focus on the stochastic
block model as the basic generative process, from which efficient
nonparametric inference can be performed, and yields a principled method
to infer hierarchical community structure from noisy data. We
demonstrate the efficacy of our approach with a variety of empirical and
artificial networks.
\end{abstract}

\maketitle

\section{Introduction}

The study of network systems of various kinds constitutes a significant
fraction of contemporary interdisciplinary research in physics, biology,
computer science and social sciences, among other
disciplines~\cite{barabasi_network_2011}. This is motivated in large
part by the surging availability of network data during the past couple
of decades, which describe the detailed interactions among constituents
of large-scale complex systems, such as transportation networks, cell
metabolism, social contacts, the internet, and various others. Despite
the widespread growth of this field, its relative infancy is still
noticeable in some aspects. In particular, even though sophisticated and
successful models of network structure and function have been proposed,
as well as powerful data analysis methods, most studies of empirical
data are performed without taking into account measurement error. Most
typically, real networks are represented as adjacency matrices,
sometimes enriched with additional information such as edge weights and
types, as well as various kinds of node properties, the validity of
which is simply taken for granted. But as is true for any empirical
scenario, network data is subject to observational errors: parts of the
network might not have been recorded, and the parts that have might be
wrong. Although this problem has been recognized in the past in several
studies~\cite{marsden_network_1990,butts_network_2003,
liben-nowell_link-prediction_2007,clauset_hierarchical_2008,
guimera_missing_2009,lu_link_2011,znidarsic_non-response_2012,
squartini_maximum-entropy_2017, newman_network_2018,
newman_network_2018-1}, the practice of ignoring measurement error is
still mainstream, and robust methods to take it into account are
underdeveloped. This is in no small part due to the fact that most
available network data contain no quantitative error assessment
information of any kind, thus preventing primary experimental
uncertainties to be propagated up the chain of analysis.

In this work we formulate a principled method to reconstruct networks
that have been imperfectly measured. We do so by simultaneously
formulating generative models of network structure --- that incorporate
degree heterogeneity, modules and hierarchies --- as well as
models of the noisy measurement process. By performing Bayesian
statistical inference of this joint model, we are able to reconstruct
the underlying network given an imperfect measurement affected by
observational noise. Importantly, our method works also when a single
measurement of the underlying network has been made, and the noise
magnitudes are unknown. This means it can be directly applied to the
majority of network data without available error estimates. In addition
to this, our method is capable of extracting hierarchical modular
structure from such noisy networks, thus generalizing the task of
community detection to this uncertain setting.

Our method is equally applicable when information on measurement error
is available, either as repeated measurements or as estimated edge
probabilities. For this class of data, we construct a general model that
allows for heterogeneous errors, that vary in different parts of the
network. We show strong empirical evidence for the existence of this
kind of heterogeneity, and demonstrate the efficacy of our method to
include it in the reconstruction.

Our method shares some underlying similarities with well known
model-based approaches of edge
prediction~\cite{clauset_hierarchical_2008,guimera_missing_2009}, but is
different from them in fundamental aspects. Most importantly,
model-based edge prediction methods yield \emph{relative} probabilities
of edges existing or not, given a generative model fitted to the
observed data. These relative probabilities can be used to reconstruct a
network provided one knows how many edges are missing or spurious. Our
method obviates the need for this information (which is in general
unknown), and yields not only a reconstructed network, but also the
uncertainty estimate that must come with it, via a posterior
distribution over all possible reconstructions. Thus our method realizes
the underlying promise of reconstruction that motivates most edge
prediction methods, but in a principled and nonparametric way.

We form the basis of our reconstruction scenario on
Ref.~\cite{newman_network_2018}, which defined a statistical inference
method based on multiple measurements of network data, but here we use a
different approach based on nonparametric Bayesian inference, combined
with community detection. This yields a more powerful method that,
differently from Ref.~\cite{newman_network_2018}, can be applied also
when the network data does not contains any kind of primary error
estimate, such as when the edges and nonedges have been measured only
once.

This work is organized as follows. In Sec.~\ref{sec:reconstruction} we
formulate our Bayesian reconstruction framework. In
Sec.~\ref{sec:measurement} we present our measurement model, and in
Sec.~\ref{sec:examples} we illustrate the use of our reconstruction
method with some examples. In Sec.~\ref{sec:performance} we perform a
detailed analysis of the reconstruction performance of the method, as
well as its use to provide estimates of various network properties. In
Sec.~\ref{sec:empirical} we employ our approach to some empirical
network data without primary error estimates, and evaluate their
reliability. In Sec.~\ref{sec:heterogeneous} we extend our method to
heterogeneous errors, and use it to analyze network data with multiple
measurements. In Sec.~\ref{sec:foreign} we show how our method can be
extended to situations where the arbitrary error estimates are
extrinsically provided, and we finalize in Sec.~\ref{sec:conclusion}
with a conclusion.

\section{Bayesian network reconstruction}\label{sec:reconstruction}

The scenario we consider is one where instead of a direct observation of
a network $\A$, we perform a noisy measurement $\D$ that contains only
indirect information about $\A$. The task of network reconstruction is
then to obtain $\A$ from $\D$. The approach we take is to perform
statistical inference, where first we model the network generating
process via a probability
\begin{equation}
  P(\A|\theta),
\end{equation}
where $\theta$ are arbitrary model parameters. The entire data
generating process is then completed by modelling also the noisy
measurement,
\begin{equation}
  P(\D|\A,\phi),
\end{equation}
conditioned on the generated network $\A$ (the ``true'' network) and
some further parameters $\phi$. Given this general setup, the
reconstruction procedure consists of determining $\A$ from the posterior
distribution
\begin{equation}\label{eq:post_general}
  P(\A|\D) = \frac{P(\D|\A)P(\A)}{P(\D)},
\end{equation}
where
\begin{equation}
  P(\D|\A) = \int P(\D|\A,\phi)P(\phi)\;\dd\phi,
\end{equation}
is the marginal probability of the measurements $\D$, and
\begin{equation}
  P(\A) = \int P(\A|\theta)P(\theta)\;\dd\theta,
\end{equation}
is the prior probability for $\A$, summed over all possible parameter
choices, weighted according to their (hyper-)prior probabilities. The remaining
term $P(\D)=\sum_{\A}P(\D|\A)P(\A)$ is a normalization constant that
corresponds to the total probability --- or \emph{evidence} --- for the
observed measurement. In the above, the probabilities $P(\theta)$ and
$P(\phi)$ encode our prior knowledge (or lack thereof) about the network
generation and measurement processes, respectively. With these at hand,
Eq.~\ref{eq:post_general} assigns the probability of a given network
$\A$ being responsible for measurement $\D$. Importantly, this
distribution defines an ensemble of possibilities for the underlying
network $\A$ that incorporates the amount uncertainty resulting from the
measurement. This contrasts with reconstruction approaches that attempt
to reproduce a single network, although within the above framework we
could also attempt to find the single most likely reconstruction that
maximizes Eq.~\ref{eq:post_general}, i.e. \emph{a maximum posterior
point estimate}. However, as we will see below, this is not the most
appropriate point estimate, as it tends to incorporate noise from the
data, biasing the reconstruction. Instead, we should consider the
consensus of the full posterior distribution, which can also give us an
estimation of uncertainty.

The above framework is general, and can be used for any kind of
generative and measurement processes. Here, we are interested in those
that can be used to describe the large-scale modular structures of
networks, characterized by the partition of the nodes into groups
$\bb=\{b_i\}$, where $b_i\in\{1,\dots,B\}$ is group membership of node
$i$. The simplest and most commonly used model for this is the
stochastic block model (SBM)~\cite{holland_stochastic_1983},
\begin{equation}
  P(\A|\bm{\omega},\bb) = \prod_{i<j}\omega_{b_i,b_j}^{A_{ij}}(1-\omega_{b_i,b_j})^{1-A_{ij}}
\end{equation}
where $\omega_{rs}$ is the probability of an edge existing between nodes
of groups $r$ and $s$. Alternatively, we could also consider a more
realistic version called the degree-corrected SBM
(DC-SBM)~\cite{karrer_stochastic_2011},
\begin{equation}
  P(\A|\bm\lambda,\bm\kappa,\bb) = \prod_{i<j}
  \frac{e^{-\kappa_i\kappa_j\lambda_{b_i,b_j}}(\kappa_i\kappa_j\lambda_{b_i,b_j})^{A_{ij}}}{A_{ij}!},
\end{equation}
where $\lambda_{rs}$ controls the number of edges between groups $r$ and
$s$ and $\kappa_i$ the expected degree of node $i$. This model variant
decouples the degrees from the group memberships, allowing for arbitrary
degree variability inside modules, a feature often found to be more
compatible with real networks~\cite{peixoto_nonparametric_2017}. (Note
that the DC-SBM generates multigraphs with $A_{ij}\in \mathbb{N}$,
whereas the SBM above generates simple graphs with $A_{ij}\in\{0,1\}$,
as our framework requires. In appendix~\ref{app:multigraphs} we amend
this inconsistency.)  Using the above, we compute the marginal network
probability as
\begin{equation}\label{sbm:evidence}
  P(\A) = \sum_bP(\A|\bb)P(\bb),
\end{equation}
with
\begin{equation}\label{sbm:marg}
  P(\A|\bb) =\int P(\A|\bm\lambda,\bm\kappa,\bb)P(\bm\kappa|\bb)P(\bm\lambda|\bb)\;\dd\kappa\,\dd\lambda,
\end{equation}
integrated over the remaining model parameters, weighted by their
respective prior probabilities. However, although Eq.~\ref{sbm:marg} can
be computed exactly~\cite{peixoto_nonparametric_2017}, the complete
marginal of Eq.~\ref{sbm:evidence} cannot, as it involves an intractable
sum over all possible network partitions. Hence, instead of computing
directly the posterior of Eq.~\ref{eq:post_general}, we obtain the joint
posterior\footnote{It is important to distinguish between the network
generation given by the prior of Eq.~\ref{sbm:marg} and the
reconstruction given by the posterior of
Eq.~\ref{eq:joint_posterior}. The former is a generative process that,
even if it closely captures the large-scale structure present in the
underlying network, it may deviate from it in important ways,
e.g. lack an abundance of triangles or other properties not well
described by the SBM, and thus generates the true network with
only a very small probability. In contrast, the posterior of
Eq.~\ref{eq:joint_posterior} corresponds to a distribution of networks
that are ``centered'' around the observed data, and will incorporate
features that are present in it, even if they are not well described by
the SBM prior (such as clustering, and other ``small-scale''
properties).}
\begin{equation}\label{eq:joint_posterior}
  P(\A,\bb|\D) = \frac{P(\D|\A)P(\A|\bb)P(\bb)}{P(\D)},
\end{equation}
which involves only quantities that can be computed exactly, except
$P(\D)$, which as we will shortly see, is unnecessary for the inference
procedure. We do the above without any loss, as the original posterior
of Eq.~\ref{eq:post_general} can be obtained by marginalization, i.e.
\begin{equation}
  P(\A|\D) = \sum_{\bb} P(\A,\bb|\D).
\end{equation}
This means that if we can sample from the joint posterior
$P(\A,\bb|\D)$, we can compute any estimate $\hat{y}$ of a network
property $y(\A)$ (e.g. the clustering coefficient) over the full
marginal $P(\A|\D)$ by averaging it over the joint posterior, i.e.
\begin{equation}
  \hat{y} = \sum_{\A}y(\A)P(\A|\D) = \sum_{\A,\bb}y(\A)P(\A,\bb|\D).
\end{equation}
The procedure we use to sample from the posterior distribution is Markov
chain Monte Carlo (MCMC). We consider move proposals of the kind
$P(\bb'|\A,\bb)$ and $P(\A'|\A,\bb)$ for the partition and network,
respectively, and accept the proposal according to the
Metropolis-Hastings~\cite{metropolis_equation_1953,hastings_monte_1970}
probability
\begin{equation}\label{eq:metropolis}
  \min\left(1, \frac{P(\A',\bb'|\D)P(\A|\A',\bb')P(\bb|\A',\bb')}{P(\A,\bb|\D)P(\A'|\A,\bb)P(\bb'|\A,\bb)}\right),
\end{equation}
which enforces detailed balance. If the move proposals are ergodic,
i.e. they allow every network $\A$ and partition $\bb$ to be proposed
eventually, this algorithm will generate samples from the posterior
distribution $P(\A,\bb|\D)$ after a sufficiently large number of
iterations (usually determined by requiring that statistical properties
of the chain, such as average log-probability, become stationary). The
ratio in Eq.~\ref{eq:metropolis} can be determined exactly without
computing the intractable constant $P(\D)$ in
Eq.~\ref{eq:joint_posterior}, making this method asymptotically
exact. We give more technical details of our MCMC procedure in
Appendix~\ref{app:mcmc}.

The above setup is still sufficiently general that it can be used with
any variant of the SBM. In particular, here we will make extensive use
of the hierarchical DC-SBM
(HDC-SBM)~\cite{peixoto_hierarchical_2014,peixoto_nonparametric_2017},
which differs from the DC-SBM in that a nested hierarchy of priors and
hyperpriors is used in place of the single prior $P(\bm\lambda|\bb)$ for
the connections between groups. In this model, groups are clustered
hierarchically into meta-groups, yielding a nested hierarchical
partition $\{\bb^l\}$, where $\bb^l$ is the partition of the groups in
level $l$. As discussed in
Refs.~\cite{peixoto_hierarchical_2014,peixoto_nonparametric_2017}, this
choice of structured priors removes a tendency of noninformative priors
to underfit~\cite{peixoto_parsimonious_2013}, and enables the detection
of structures at multiple scales, while at the same time remaining
unbiased with respect to different types of mixing patterns. Its
posterior distribution is obtained in the same fashion, following the
framework above, simply by replacing $\bb\to\{\bb^l\}$.

In the following, whenever we mention that we sample from the posterior
$P(\A|\D)$, it is meant we sample from the joint posterior
$P(\A,\bb|\D)$, and marginalize over $\bb$, as described above. The same
is true when using the hierarchical model, i.e. we sample from
$P(\A,\{\bb^l\}|\D)$, and marginalize over the hierarchical partitions
$\{\bb^l\}$.

The main difference from typical community detection based on
statistical inference is that here we are not only interested in
detecting modules in networks, but also inferring the network
itself. Therefore, both the network and its partition into
(hierarchical) groups are inferred from indirect data. As we will see,
the simultaneous detection of modules offers a substantial advantage to
the reconstruction task, as it allows correlations among edges to inform
it. This means that we are able to perform reconstruction in situations
which would otherwise be impossible. But before we proceed, we need to
model the measurement process itself, as we do in the following.

\subsection{Noisy network measurements}\label{sec:measurement}

Here we will consider the scenario used in
Ref.~\cite{newman_network_2018}, where the edges of a network are
measured directly and repeatedly, but the process is noisy, and
potentially distorts the network. In particular, we will assume that for
each node pair $(i,j)$ we perform $n_{ij}$ distinct measurements, and
record $x_{ij}$ positive outcomes,
i.e. an edge is observed. For each observation, we have a probability
$p$ of observing a missing edge (i.e. a false negative) and a
probability $q$ of observing a spurious edge (i.e. a false positive),
depending in each case if the underlying network possesses or not an
edge $(i,j)$. Thus, for each edge the observation probability is
distributed according to a binomial distribution, with a success rate
that depends on whether an edge exists in the underlying network, i.e.
\begin{multline}
  P(x_{ij}|n_{ij}, A_{ij}, p, q) = \\
  {n_{ij}\choose
  x_{ij}}\left[(1-p)^{x_{ij}}p^{n_{ij}-x_{ij}}\right]^{A_{ij}}\left[q^{x_{ij}}(1-q)^{n_{ij}-x_{ij}}\right]^{1-A_{ij}}.
\end{multline}
Thus, the joint likelihood for the whole set of measurements $\x=\{x_{ij}\}$ is
\begin{multline}\label{eq:mlikelihood}
  P(\x|\n,\A,p,q)= \prod_{i<j}P(x_{ij}|n_{ij}, A_{ij}, p, q)\\
    =\left[\prod_{i<j}{n_{ij}\choose x_{ij}}\right](1-p)^{\T}p^{\E-\T}q^{\X-\T}(1-q)^{\M-\X-\E+\T},
\end{multline}
written in terms of the following summary quantities,
\begin{align}
  \M&=\sum_{i<j}n_{ij},  &  \X&=\sum_{i<j}x_{ij},\\
  \E&=\sum_{i<j}n_{ij}A_{ij}, &  \T&=\sum_{i<j}x_{ij}A_{ij},
\end{align}
where $\M$ is the total number of measurements (edge or nonedge), $\X$
is the total number of observed edges, $\E$ is the total number of
measured edges and $\T$ is the total number of correctly observed
edges.\footnote{Note that the binomial terms in
Eq.~\ref{eq:mlikelihood}, and those that follow it, only depend on the
measurement data, not on $\A$, $p$ or $q$, so ultimately they will not
contribute to the posterior distribution.} From this, we also identify
the total number of false positives (spurious edges) as $\X-\T$ and of
false negatives (missing edges) as $\E-\T$.

To proceed with our calculation we need to specify the degree of prior
knowledge we have on the error rates $p$ and $q$. We can express this
most naturally with a Beta distribution,
\begin{equation}\label{eq:beta}
  P(p|\alpha,\beta) = \frac{p^{\alpha-1}(1-p)^{\beta-1}}{\mathcal{B}(\alpha,\beta)}
\end{equation}
where $\mathcal{B}(x,y)=\Gamma(x)\Gamma(y)/\Gamma(x+y)$ is the Euler
beta function, and $\Gamma(x)$ is the gamma function, and likewise for
$P(q|\mu,\nu)$, with hyperparameters $\mu$ and $\nu$. As illustrated in
Fig.~\ref{fig:beta} of appendix~\ref{app:beta}, a value of
$\alpha=\beta=1$ encodes a maximum amount of prior ignorance with
respect to $p$, which is then uniformly distributed in the unit
interval. Conversely, values $\alpha\to\infty$ and $\beta\to\infty$
converge to a Dirac delta function centered at $\alpha/(\alpha+\beta)$,
amounting to a maximum certainty for a particular value of $p$, and
therefore intermediary values of $\alpha$ and $\beta$ interpolate
between these two extremes (and analogously for $q$ with $\mu$ and
$\nu$). With this, we can compute the integrated likelihood
\begin{multline}
  P(\x|\n,\A,\alpha,\beta,\mu,\nu)\\
  = \int P(\x|\n,\A,p,q)P(p|\alpha,\beta)P(q|\mu,\nu)\;\dd p\,\dd q\\
  =\left[\prod_{i<j}{n_{ij}\choose x_{ij}}\right]
  \frac{\mathcal{B}(\E-\T+\alpha, \T+\beta)}{\mathcal{}B(\alpha,\beta)}\times\\
  \frac{\mathcal{B}(\X-\T+\mu, \M-\X-\E+\T+\nu)}{\mathcal{B}(\mu,\nu)}.
\end{multline}
The noninformative case $\alpha=\beta=\mu=\nu=1$ simplifies further to
\begin{multline}\label{eq:xmarg}
  P(\x|\n,\A) =\left[\prod_{i<j}{n_{ij}\choose x_{ij}}\right]\times\\
  {\E \choose \T}^{-1}\frac{1}{\E+1}{\M-\E \choose \X-\T}^{-1}\frac{1}{\M-\E+1}.
\end{multline}
The above noninformative generative process can also be equivalently
interpreted as first choosing the number of false positives $\X-\T$
uniformly from the interval $[0,\M-\E]$ and then selecting them
uniformly at random from the possible set with ${\M-\E\choose \X-\T}$
elements, and similarly choosing the number of false-negatives $\E-\T$
uniformly in the interval $[0, \E]$ and the false-negatives from the set
of size ${\E\choose \E-\T} = {\E\choose \T}$.

With the integrated likelihood in place, we can finally complete the
posterior distribution of Eq.~\ref{eq:post_general} with $\D=(\n,\x)$,
which in this case becomes,
\begin{equation}\label{eq:measured_posterior}
  P(\A|\n,\x,\alpha,\beta,\mu,\nu) = \frac{P(\x|\n,\A,\alpha,\beta,\mu,\nu)P(\A)}{P(\x|\alpha,\beta,\mu,\nu)}.
\end{equation}
For $P(\A)$ we will use the SBM and sample $\A$ using MCMC from the
joint posterior $P(\A,\bb|\n,\x,\alpha,\beta,\mu,\nu)$, as discussed
previously.

Even though we have integrated over the error probabilities $p$ and $q$
in the above, we can nevertheless obtain their posterior estimates by
averaging from the above posterior
\begin{multline}
  P(p|\n,\x,\alpha,\beta,\mu,\nu)=\\
  \sum_{\A}P(p|\n,\x,\A,\alpha,\beta)P(\A|\n,\x,\alpha,\beta,\mu,\nu),
\end{multline}
using the posterior for $p$ conditioned on the network $\A$,
\begin{equation}
  P(p|\n,\x,\A,\alpha,\beta) = \frac{p^{\E-\T+\alpha-1}(1-p)^{\T+\beta-1}}{\mathcal{B}(\E-\T+\alpha, \T+\beta)}
\end{equation}
and likewise for $q$ with
\begin{equation}
  P(q|\n,\x,\A,\mu,\nu) = \frac{q^{\X-\T+\mu-1}(1-q)^{\M-\X-\E+\T+\mu-1}}{\mathcal{B}(\X-\T+\mu, \M-\X-\E+\T+\nu)}.
\end{equation}

In the following, we will most often assume the noninformative case
$\alpha=\beta=\nu=\mu=1$, corresponding to the maximum lack of prior
knowledge about the measurement noise. In order to unclutter our
expressions, if this is the case we will simply omit those
hyperparameters from the posterior distribution,
i.e. $P(\A|\n,\x) \equiv P(\A|\n,\x,\alpha=1,\beta=1,\mu=1,\nu=1)$.

\subsubsection{Single edge measurements}\label{sec:single}

As we increase the number of measurements $n_{ij}$ of each pair of
nodes, we should expect also to increase the accuracy of the
reconstruction, resulting in a posterior distribution $P(\A|\n,\x)$ that
is very sharply peaked around the true underlying network. Although this
scenario is plausible, and indeed desirable under controlled
experimental conditions, this is not representative of the majority of
the network data that are currently available. In fact, inspecting
comprehensive network catalogs such as
KONECT~\cite{kunegis_konect:_2013} and ICON~\cite{clauset_colorado_2016}
reveals a very pauper set of network data that can be cast under this
setting of repeated measurements. On the contrary, the vast majority of
them offer only a single adjacency matrix without quantitative error
estimates of any kind. Needless to say, this is no reason to assume that
they do not, in fact, contain errors, only that they have not been
assessed or published.

Here we propose an approach of assessing the uncertainty of this
dominating kind of network data by interpreting it as a single
measurement with unknown errors rates, using the framework outlined
above. In more detail, we assume that $n_{ij}=1$ for every pair $i,j$
and that the single measurements $x_{ij}\in\{0,1\}$, correspond to the
reported adjacency matrix. The lack of knowledge about the underlying
error rates $p$ and $q$ can be expressed by choosing
$\alpha=\beta=\mu=\nu=1$, in which case it is assumed that they both lie
\emph{a priori} anywhere in the unit interval.\footnote{One could argue
that being totally agnostic about the error rates $p$ and $q$ is too
extreme, as in many cases they are likely to be small in some sense,
even if we cannot precisely quantify how small at first. The answer to
this objection is that, to the extent that this vague belief can be
quantified, it should be done so via the hyperparameters
$\alpha,\beta,\gamma$ and $\mu$ --- as it can with our method ---
otherwise we have little choice but to assume maximum ignorance.} At
first we may wonder if this approach has any chance of succeeding, since
the lack of knowledge about the error rates means that the network could
have been modified in arbitrary ways, such that the true underlying
network is radically different from what has been observed. Indeed, if
we define the distance between measured and generated networks,
\begin{equation}\label{eq:distance}
  d(\A,\x) = \sum_{i<j}\left|A_{ij}-x_{ij}\right| = (\E-\T) + (\X-\T),
\end{equation}
which equals the sum of false negatives and false positives, we have
that according to Eq.~\ref{eq:xmarg}, the expected distance over many
measurements is
\begin{equation}
  \bar{d}(\A) = \sum_{\x}d(\A,\x)P(\x|\n,\A) = {N\choose 2}/2,
\end{equation}
which is half the maximum possible distance of ${N\choose 2}$, which
might lead us to conclude that our noise model will invariably destroy
the network beyond the possibility of reconstruction, regardless of its
original structure. What changes this picture is the fact that the
posterior distribution $P(\A|\x,\n)$ of Eq.~\ref{eq:measured_posterior}
will in fact be more concentrated on the generated network than the
implied by the above, and ultimately will depend crucially on our
generative process $P(\A)$. The first point can be made by assuming a
fully random generative model,
\begin{equation}
  P(\A|\omega) = \prod_{i<j}\omega^{A_{ij}}(1-\omega)^{1-A_{ij}},
\end{equation}
which means that the true networks being measured are assumed to be
completely random, given a particular density $\omega$. The full prior
can be obtained by a noninformative assumption $P(\omega)=1$, which
yields
\begin{align}\label{eq:er_prior}
  P(\A) &= \int P(\A|\omega)P(\omega)\,\dd\omega,\\
   &={{N\choose 2}\choose E}^{-1}\frac{1}{{N\choose 2}+1},
\end{align}
with $E=\sum_{i<j}A_{ij}=\E$ being the total number of edges, which is
equivalent to sampling to the total number of edges from the interval
$[0, {N\choose 2}]$ and then a fully random graph with that number of
edges. Combining this with Eq.~\ref{eq:xmarg}, yields the posterior
distribution, which can be written as the product of two conditional
probabilities,
\begin{equation}
  P(\A|\x,\n) = P(\A|\x,\T,\E)P(\T,\E|\x),
\end{equation}
with
\begin{equation}
  P(\A|\x,\T,\E) = {\X\choose \X-\T}^{-1}{{N\choose 2}-X \choose \E - \T}^{-1}
\end{equation}
corresponding to the uniform sampling of $\A$ with exactly $\E-\T$ false
negatives and $\X-\T$ false positives, and
\begin{equation}
  P(\T,\E|\x) \propto \frac{[\T\le \E][\T\le \X]}{(\E+1)[{N\choose 2}-\E+1]}
\end{equation}
with $[\cdots]$ being the Inverson bracket that equals $1$ if the
condition inside it is true, or $0$ otherwise, determines the posterior
probability of the number of false negatives and false positives, up to
a normalization constant. Although this distribution decays for values
of $\E$ larger than $0$, the decay is slow with $\sim 1/\E$, and hence,
on average, the inferred networks $\A$ sampled from $ P(\A|\x,\n)$ will
be dense, yielding large distances $d(\A,\A^*)$ if the true generated
network $\A^*$ is sparse. Although the posterior distribution of false
negatives and positives resulting from $P(\T,\E|\x)$ is not uniformly
distributed in the allowed interval, it is also not sufficiently
concentrated to enable any reasonable accuracy in the reconstruction,
regardless of how large the network is. What changes this considerably
is to replace the fully random model of Eq.~\ref{eq:er_prior} by a more
structured model. The key observation here is that the modifications
induced by the error rates $p$ and $q$ affect uniformly every edge and
nonedge, and thus with structured models we can exploit the observed
correlations in the measurements $\x$ to infer the underlying network
$\A$, and in fact even the error rates $p$ and $q$, which are \emph{a
priori} unknown.

\begin{figure}
  \includegraphics{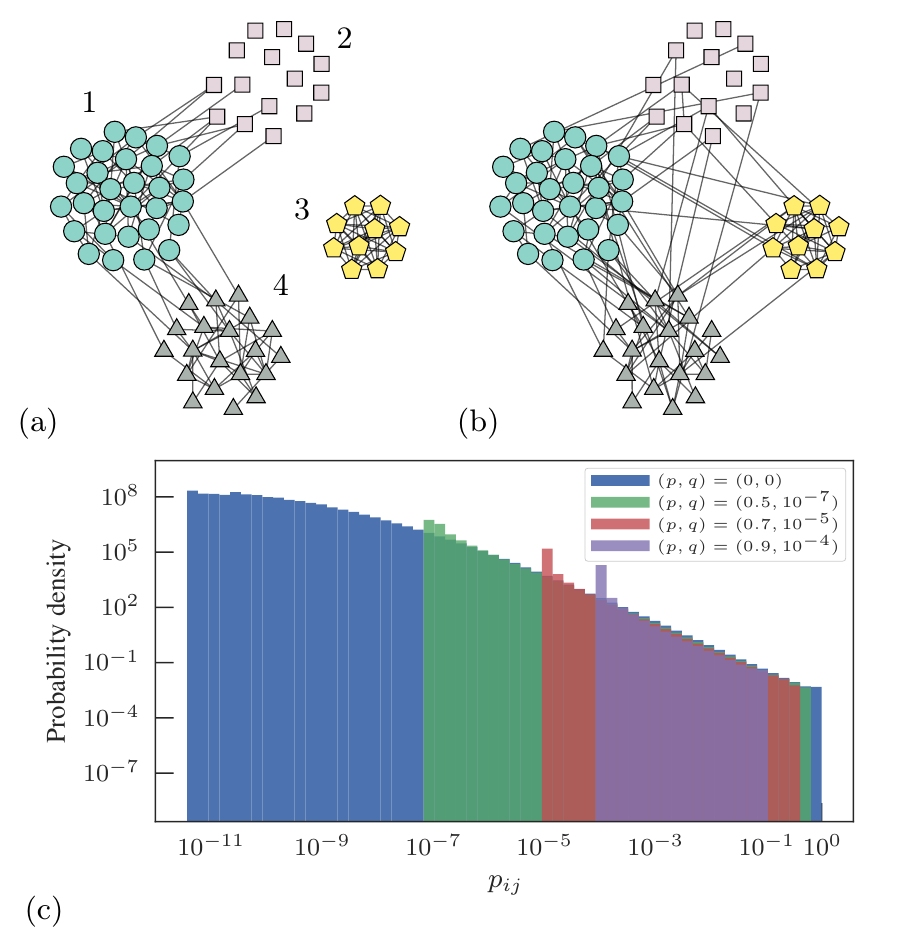}

  \caption{(a) Illustration of a hypothetical measured network, with
  \emph{a priori} unknown errors, but from which error estimates can be
  made: the lack of edges between groups 2 and 3, 3 and 4, 2 and 4, and
  1 and 3 implies that the probability $q$ of missing edges is likely to
  be low. Similarly, the large internal density of group 3 (which forms
  a clique of 10 nodes) implies that the missing edge probability $p$
  must be low as well. (b) How the network in (a) would look like for
  higher values of $p$ and $q$. (c) The distribution of marginal edge
  probabilities $p_{ij}$ between every node pair, for a fit of the
  HDC-SBM on the openflights data (see Appendix~\ref{app:data}),
  measured with different values of the noise parameters $(p,q)$. As the
  noise magnitudes increase, the probabilities become less
  heterogeneous, and concentrate in narrower intervals. Hence, the
  inference of broad connection probabilities from data rules out the
  existence of strong noise in the measurement.\label{fig:densities}}
\end{figure}

We illustrate this by considering the non-degree-corrected SBM, where
networks are generated with probability
\begin{equation}
  P(\A|\bm\omega,\bb) = \prod_{i<j}\omega_{b_ib_j}^{A_{ij}}(1-\omega_{b_ib_j})^{1-A_{ij}}.
\end{equation}
The final likelihood for the measurements $\x$ in this case will be
identical to an effective SBM, given by
\begin{align}
  P(\x|\n,p,q,\bm\omega,\bb) &= \sum_{\A}P(\x|\n,\A,p,q)P(\A|\bm\omega,\bb)\\
  &=\prod_{i<j}{\omega_{b_ib_j}'}^{x_{ij}}(1-\omega_{b_ib_j}')^{1-x_{ij}}
\end{align}
where
\begin{equation}\label{eq:sbm_eff}
  \omega_{rs}' = (1-p-q)\omega_{rs}+q
\end{equation}
are the new effective SBM probabilities that have been scaled and
shifted by the measurement noise. Suppose, for simplicity, that we know
the true network partition $\bb$, and that the number of groups is very
small compared to the number of nodes in each group. In this situation,
the posterior distribution for $\bm\omega'$ should be tightly peaked
around the maximum likelihood estimate $\bm{\hat\omega'}$,
\begin{equation}
  \hat\omega'_{rs} = (1-p-q)\omega_{rs}+q = \frac{e_{rs}}{n_rn_s},
\end{equation}
where $e_{rs}=\sum_{ij}x_{ij}\delta_{b_i,r}\delta_{b_j,s}$ is the number
of observed edges between groups $r$ and $s$ (or twice that for $r=s$)
and $n_r$ is the number of nodes in group $r$. The joint posterior
distribution for $p$ and $q$ will then be asymptotically given by
\begin{multline}
  P(p,q|\x,\n,\bb) \propto \int P(\x|\n,p,q,\bm\omega,\bb)P(\bm\omega|\bb)\;\dd\bm\omega\\
  \begin{aligned}
  &\propto \prod_{r\le s}\int_0^1\delta((1-p-q)\omega_{rs}+q-e_{rs}/n_rn_s)P(\omega_{rs}|\bb)\;\dd\omega_{rs}\\
  &\propto \prod_{r\le s}\left[0\le\frac{e_{rs}/n_rn_s-q}{1-p-q}\le 1\right]\frac{P\left(\frac{e_{rs}/n_rn_s-q}{1-p-q}\middle|\bb\right)}{1-p-q},
  \end{aligned}
\end{multline}
up to normalization, where $[\cdots]$ is again the Inverson bracket. The
constraints above imply that the inferred error rates will be bounded by
the maximum and minimum inferred connection probabilities, i.e.
\begin{align}\label{eq:bounds}
  \hat q &\le \min_{rs} \frac{e_{rs}}{n_rn_s},\\
  \hat p &\le 1 - \max_{rs} \frac{e_{rs}}{n_rn_s}.
\end{align}
These bounds mean that if we have not observed many edges between groups
$r$ and $s$, this implies that $q$ could not have been very large. If
instead we do observe many edges between these groups, then this means
that the value of $p$ could not have been very large either (see
Fig.~\ref{fig:densities}a and b). This holds for every pair of groups
$r$ and $s$, but the values of $p$ and $q$ are global. Therefore, as
long as the inferred SBM probabilities are sufficiently
\emph{heterogeneous}, they should constrain the inferred error rates to
narrow intervals --- which will also constrain the inferred number of
false negatives and false positives (see
Fig.~\ref{fig:densities}c).\footnote{We stress that the bounds of
Eq.~\ref{eq:bounds} are strict only in the limit of dense network with
few groups, and do not represent the posterior distribution found for
arbitrary data. These bounds are presented just to convey the intuition
of how structure heterogeneity can inform the error probabilities.} On
the other hand, if the model probabilities are homogeneous, the
posterior distribution for the errors will be broad, and the quality of
the reconstruction will be poor. Therefore, the success of this approach
depends ultimately on the observed networks being sufficiently
structured, and of our models being capable of describing them.

\begin{figure*}
  \includegraphics{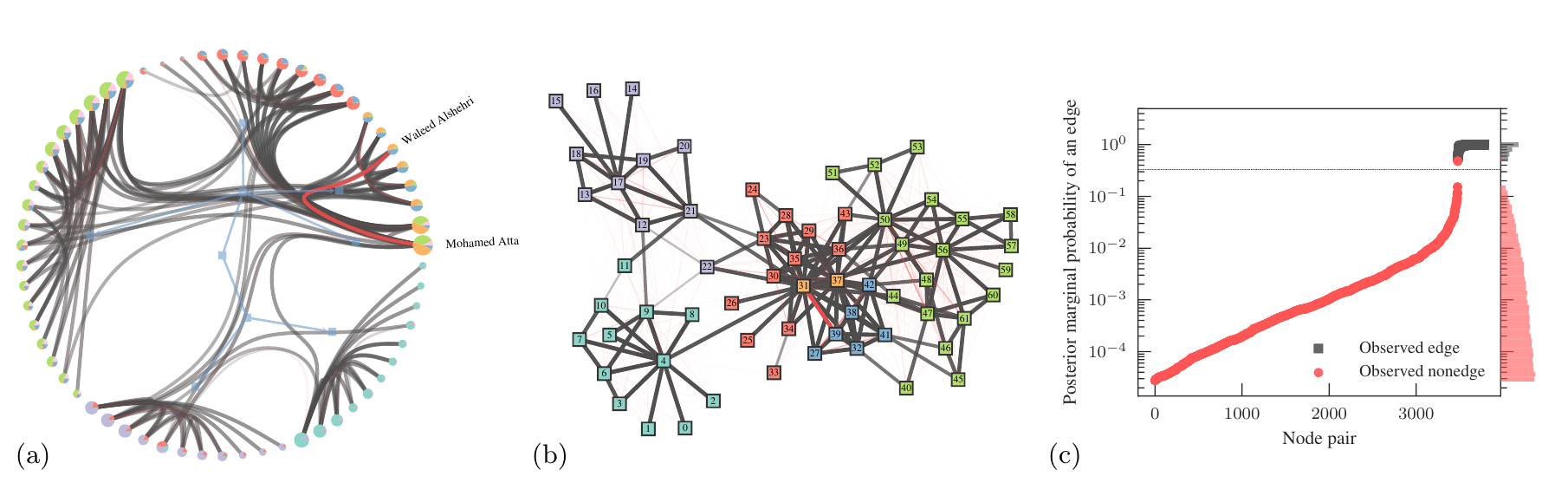}

  \caption{Network of social associations between 9/11
  terrorists~\cite{krebs_mapping_2002,krebs_uncloaking_2002}. This
  network was measured by potentially unreliable means, but no
  quantitative error estimates are known, and no repeated measurements
  were made. In (a) and
  (b) is shown the inferred network according to our method --- which
  does not require direct error estimates or repeated measurements ---
  where the edge thickness indicates the posterior marginal probability
  of an edge existing. In
  (a) the inferred hierarchical structure is shown, with pie charts on
  the nodes indicating the marginal probabilities of group memberships,
  and in (b) a spatial layout of the same network shows the lowest level
  of the hierarchy as the node colors. The edge shown in red is inferred
  as existing with a large probability, despite not being
  measured. Other potentially missing edges are also shown in red, with
  a probability given by their thickness and opacity. In (c) is shown
  the marginal probability of edge existence for all node pairs,
  indicating a fair amount of inferred reliability --- with the
  exception of the single missing edge highlighted in (a) and (b) ---
  despite the lack of direct error estimates in the data. The horizontal
  line marks a 1/3 probability as a visual aid. The missing edge
  corresponds to a connection between Mohamed Atta and Waleed Alshehri,
  which was not considered in
  Refs.~\cite{krebs_mapping_2002,krebs_uncloaking_2002}, but is
  corroborated by reports that they shared an apartment in Berlin, and
  met previously in Spain.
  \label{fig:terrorists}
  }
\end{figure*}

The above means that we have a better chance of accurate reconstruction
if our models are capable of detecting heterogeneous connection
probabilities among nodes. A fully uniform model like the
Erd\H{o}s-Renyi of Eq.~\ref{eq:er_prior} (equivalent to a SBM with only
one group) will exhibit the worse possible performance. The DC-SBM, on
the other hand, should in general perform better than the SBM, since it
is capable of capturing degree heterogeneity inside groups, which is a
common feature of many
networks~\cite{karrer_stochastic_2011,peixoto_nonparametric_2017}. The
HDC-SBM~\cite{peixoto_hierarchical_2014, peixoto_nonparametric_2017}
should perform even better, since its tendency not to underfit means it
can detect statistically significant structures at smaller scales.

Finally, it must also be noted that when performing only single
measurements, there remains an unavoidable identification problem, where
it becomes impossible to fully distinguish a network that has been
sampled from a SBM with parameters $\bm\omega$ and error rates $p$ and
$q$ from the same network sampled from a SBM with parameters
$\bm\omega'$ given by Eq.~\ref{eq:sbm_eff} and error rates $p=q=0$ (and
in fact any interpolation between these two extremes). This uncertainty,
however, will be reflected in the variance of the posterior
distribution, and serves as a worse-case estimation of the error rates,
which ultimately can be improved either by incorporating better prior
knowledge (e.g. via the hyperparameters $\alpha,\beta,\nu$ and $\mu$) or
performing multiple measurements.

\subsection{Empirical examples}\label{sec:examples}

Before we proceed further with a systematic analysis of our
reconstruction method, we illustrate its behavior with some empirical
data that are likely to contain errors and omissions. We begin with the
network of social associations between 62 terrorists responsible for the
9/11 attacks~\cite{krebs_mapping_2002,krebs_uncloaking_2002}. The
existence of an edge between two terrorists is established if there is
evidence they interacted directly in some way, e.g. if they attended the
same college or shared an address. Clearly, this approach is inherently
unreliable, as investigators may either fail to record evidence, or the
evidence recorded may be simply erroneous. Nevertheless, although this
potential unreliability was acknowledged in
Refs.~\cite{krebs_mapping_2002,krebs_uncloaking_2002}, is was not
assessed quantitatively, and the data presented there is a single
adjacency matrix with no error estimates. Therefore it serves as a
suitable candidate for the application of our reconstruction
method. When applied to this dataset, our approach yields the results
seen in Fig.~\ref{fig:terrorists}, which shows the marginal posterior
probability of each possible edge in the network, in addition to the
hierarchical modular structured captured by the HDC-SBM. Our method
identifies the organization into a few largely disconnected cells,
typical of terrorist groups. When ranking the potential edges according
to their marginal posterior probability, as shown in
Fig.~\ref{fig:terrorists}c, we have that all observed edges are more
likely to be true edges than any of the nonedges, indicating a fair
degree of inferred reliability. The observed nonedges have a
probability substantially smaller than the observed edges of being
edges, with the sole exception of a connection between Mohamed Atta (one
of the main leaders) and Waleed al-Shehri, which was not considered in
Refs.~\cite{krebs_mapping_2002,krebs_uncloaking_2002}, but to which our
method ascribes a reasonably high probability of $0.48$. Atta is
connected to all members of al-Shehri's group, and according to the HDC-SBM
the sole missing link between them is therefore suspicious. Indeed,
journalistic reports place both individuals occasionally sharing an
apartment in Berlin,\footnote{The Washington Post,
2001. \url{https://www.washingtonpost.com/wp-srv/nation/graphics/attack/hijackers.html}}
and meeting at least once in Spain,\footnote{ABC Eyewitness News,
2001. \url{https://web.archive.org/web/20030415011752/http://abclocal.go.com/wabc/news/WABC_092701_njconnection.html}}
prior to the attacks, which seems to corroborate our reconstruction. The
remaining observed nonedges have a probability of $0.15$ or smaller,
which should not be outright discarded, and could serve as candidates
for further investigation.

\begin{figure}
  \includegraphics{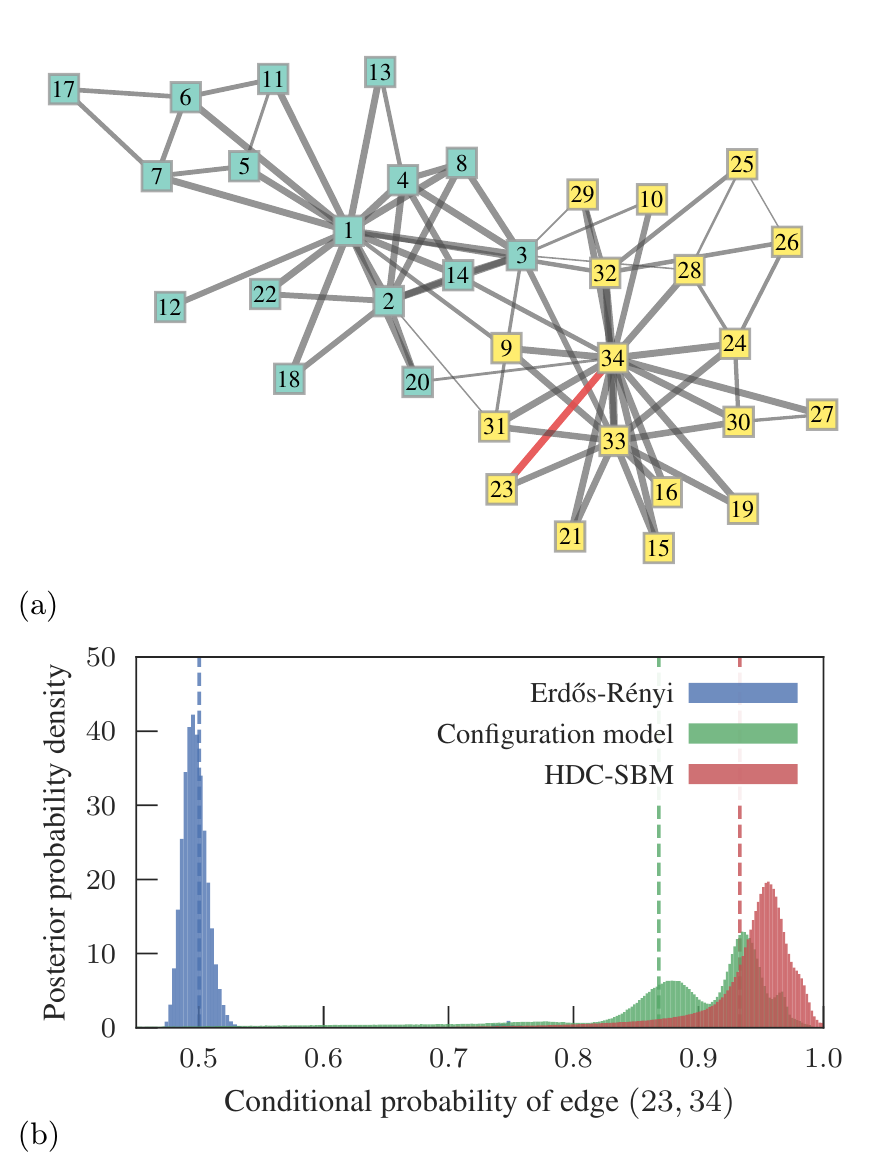}

  \caption{Inferred Zachary's karate club network using the uncertain
    data from the original publication~\cite{zachary_information_1977},
    which contains an ambiguous edge $(23,34)$, as explained in the
    text. (a) Layout of the reconstructed network showing the posterior
    edge probabilities as edge thickness, according to the HDC-SBM, and
    the ambiguous edge in red. The node colors correspond to a sample
    from the posterior distribution of the node partitions.  (b)
    Posterior probability density of the probability of edge $(23, 34)$,
    conditioned on the remaining edges and model parameters, for the
    different model variants indicated in the legend and explained in
    the text. The vertical dashed lines indicate the distribution
    averages, corresponding to the marginal posterior probability of the
    edge. \label{fig:karate}}
\end{figure}

We now move to another social network, namely the interactions between
34 members of a karate club, originally studied by
Zachary~\cite{zachary_information_1977}. This network has been widely
used to evaluate community detection methods, after its use for this
purpose in Ref.~\cite{girvan_community_2002}. It was recorded just
before the split of the club in two disjoint groups after a conflict,
and many community detection methods are capable of accurately
predicting the split by detecting communities from this
snapshot. However, not only does the original publication of
Ref.~\cite{zachary_information_1977} omits any assessment of measurement
uncertainties, but also it clearly contains one obvious error: the
adjacency matrix $\A$ published in the original study, although it is
supposed to be symmetric, contains two inconsistent entries with
$A_{ij}\ne A_{ji}$, for $(i,j)=(23,34)$, creating an ambiguity about the
existence of this particular edge.\footnote{To the best of our
knowledge, this issue was first identified by Aaron
Clauset~\cite{aaron_private_2018}, who assembled the alternative dataset
with $A_{23,34}=0$ and hence 77 edges (as opposed to the more common
variant with $A_{23,34}=1$ and 78 edges) and made it available in his
website c.a. 2015, \url{http://santafe.edu/~aaronc/data/zkcc-77.zip}.}
The authors of Ref.~\cite{girvan_community_2002} made the decision of
assuming $A_{23,34}=1$, even though there seems to be no obvious
reason to decide either way \emph{a priori}. The vast majority of other
works in the area followed suit (possibly inadvertently), thus
incorporating this potential, though arguably small, error in their
analysis.  Here we tackle this reconstruction problem by mapping the
uncertain dataset of Ref.~\cite{zachary_information_1977} to our
framework. Since each node pair $(i,j)$ was also presented reversed
$(j,i)$, we consider these as independent measurements, such that
$n_{ij}=2$ for every pair $(i,j)$. Since the measurements were
consistent for all but one pair, we have $x_{ij} = 2$ or $0$, except for
the offending entry with $x_{(23,34)}=1$. Based on this we employed our
reconstruction approach to obtain $P(\A|\n,\x)$, using as generative
processes the Erd\H{o}s-Rényi (ER) model (equivalent to a SBM with only
one group, $B=1$), the configuration model (CM) (equivalent to a DC-SBM
with $B=1$) and the HDC-SBM. As we see in Fig.~\ref{fig:karate}, the ER
model is incapable of disambiguating the data, as it cannot be used to
detect any structure in it, and ascribes a posterior probability of
$0.5$ to the uncertain edge. Both the CM and the HDC-SBM, however,
ascribe high probabilities for the edge, of $0.87$ and $0.93$,
respectively. The CM approach is able to recognize that since node 34 is
a hub in the network, an edge connecting to it more likely to occur than
not, and the HDC-SBM can further use the fact that both nodes belong to
the same group. Therefore, it seems like the choice made by the authors
of Ref.~\cite{girvan_community_2002} of assuming $A_{23,34}=1$ was
fortuitous, and the \emph{de facto} instance of this network used by the
majority of researchers is the one mostly likely to correspond to the
original study.

In the following we move to a systematic analysis of the reconstruction
method, based on empirical and simulated data.

\subsection{Reconstruction performance}\label{sec:performance}

Before we evaluate the performance of the reconstruction approach, we
must first decide how to quantify it. As a criterion of how close an
inferred network $\hat\A$ is to the true network $\A^*$ underlying the
data we will use the distance of Eq.~\ref{eq:distance},
\begin{equation*}
  d(\hat\A,\A^*) = \sum_{i<j}|\hat{A}_{ij}-A_{ij}^*|.
\end{equation*}
A successful reconstruction method should seek to find an estimate
$\hat\A$ that minimizes this distance. However, since we do not have
direct access to the true network $\A^*$, the best we can do is to
consider the average distance over the posterior distribution given the
noisy data,
\begin{align}
  \bar{d}(\hat\A) &= \sum_{\A}d(\hat\A,\A)P(\A|\x,\n)\\
  &= \sum_{i<j}|\hat{A}_{ij}-\pi_{ij}|,
\end{align}
where
\begin{equation}
  \pi_{ij} = \sum_{\A}A_{ij}P(\A|\x,\n),
\end{equation}
is the marginal posterior probability of edge $(i,j)$. If we minimize
$\bar{d}(\hat\A)$ with respect to $\hat\A$, we obtain
\begin{equation}\label{eq:mmp}
  \hat{A}_{ij} =
  \begin{cases}
    1 & \text{ if } \pi_{ij} > 1/2\\
    0 & \text{ if } \pi_{ij} < 1/2,
  \end{cases}
\end{equation}
for $\pi_{ij}\ne 1/2$. Eq.~\ref{eq:mmp} defines what is called a maximum
marginal posterior (MMP) estimator, and it leverages the consensus of
the entire posterior distribution of all possible networks for the
estimation of every edge. Operationally, it can be obtained very easily
by sampling networks from the posterior distribution, and computing how
often each edge is observed, yielding an estimate for $\bm\pi$ and hence
$\hat\A$.

Given the above criterion, we evaluate the reconstruction performance by
simulating the noisy measurement process. We do this by taking a real
network $\A^*$ (which for this purpose we are free to declare to be
error free), and obtaining a measurement $\x$ given error rates $p$ and
$q$, and measuring each edge and nonedge the same number of times
$n_{ij}=n$. We choose $p$ arbitrarily and $q=pE/[{N\choose 2} - E]$,
where $E$ is the number of edges in $\A^*$, so that the measured
networks have the same average density as $\A^*$. Given a final
measurement $\x$, we sample inferred networks $\A$ from the posterior
distribution $P(\A|\x,\n)$ and compute the MMP estimate $\hat\A$ from
the marginal distribution $\bm\pi$. The quality of the reconstruction is
then assessed according to the similarity to the true network $\A^*$,
$S(\hat\A,\A^*)\in[0,1]$, defined as
\begin{equation}\label{eq:similarity}
  S(\hat\A,\A^*) = 1-\frac{d(\hat\A,\A^*)}{\sum_{i<j}\hat A_{ij}+A_{ij}^*},
\end{equation}
where $d(\hat\A,\A^*)$ is the distance defined in
Eq.~\ref{eq:distance}. A value of $S(\hat\A,\A^*)=1$ indicates perfect
reconstruction, and $S(\hat\A,\A^*)=0$ the situation where $\hat\A$ and
$\A^*$ do not share a single edge.\footnote{Note that $S(\hat\A,\A^*)$
differs from the measure of accuracy commonly used in binary
classification tasks, defined as the fraction of entries in $\A^*$ (both
zeros and ones) that were correctly estimated in $\hat\A$, which in this
case amounts to $1-d(\hat\A,\A^*)/{N\choose 2}$. This is because we are
more typically interested in reconstructing sparse networks, where the
number of zeros (nonedges) is far larger than ones (edges), such that
$d(\hat\A,\A^*)\ll {N\choose 2}$, for all choices of sparse $\hat\A$ and
$\A^*$, causing the accuracy to approach one simply because $\hat\A$
shares most of its nonedges with $\A^*$, even if they do not have a
single edge in common. The similarity $S(\hat\A,\A^*)$ fixes this
problem by normalizing instead by the total number of edges observed in
both networks. Note, however, that a value of $S(\hat\A,\A^*)=0$ does
not imply that the distance $d(\hat\A,\A^*)$ is maximal, only that it is
large enough for both networks not to share any edge.}

\begin{figure*}
  \includegraphics{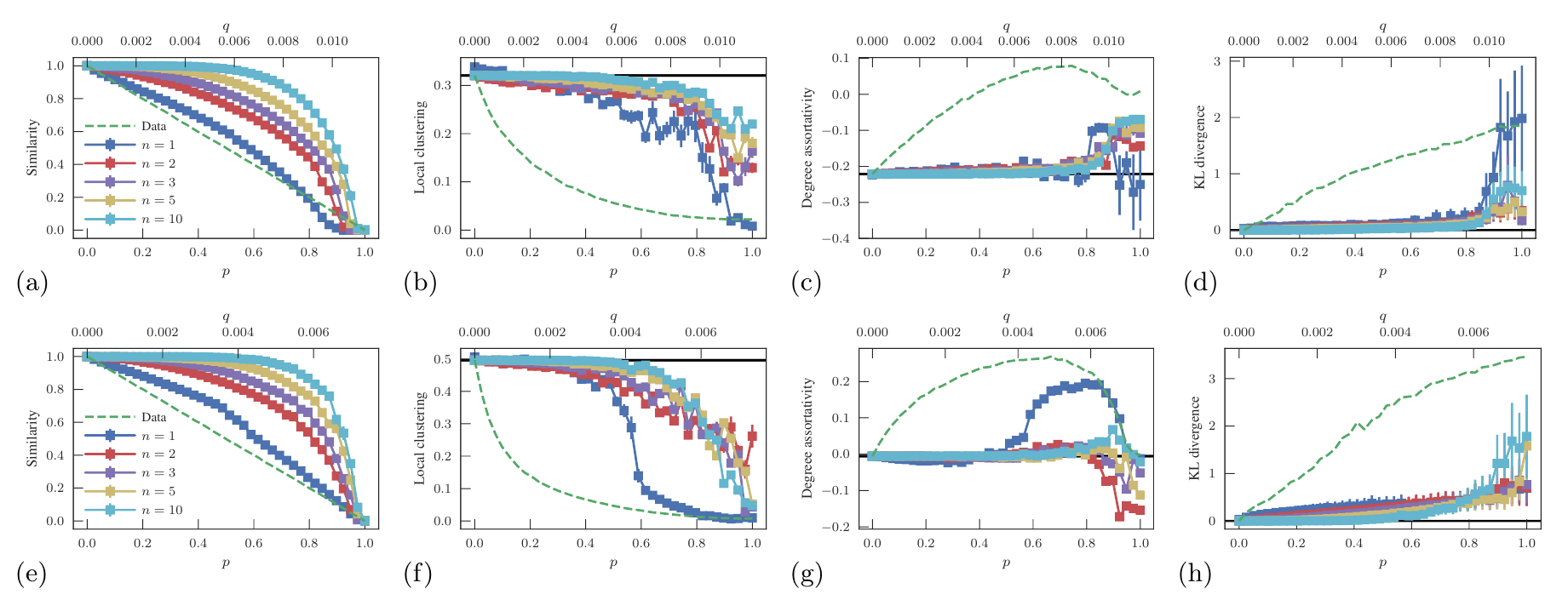}

  \caption{Reconstruction performance for political blogs
  (top row) and openflights (bottom row) networks. In each case, the
  empirical network was considered as the true network, and simulated
  measurements were made for several values of missing edge probability
  $p$, with a spurious edge probability $q=pE/[{N\choose 2} - E]$. [(a) and
  (e)] Similarity of the MMP estimator to the true network,
  $S(\hat\A,\A^*)$, as a function of $p$, and for several values of the
  number of repeated measurements, $n$. [(b), (c), (f), and (g)]
  Posterior average local clustering and degree assortativity
  coefficients, according to the same legend as (a) and (e). [(d) and
  (g)] KL divergence between true and inferred degree distributions, as
  discussed in the text.  In all cases [(a) to (h)] the dashed curve
  shows the corresponding value obtained directly with the measured data
  with $n=1$, and the solid horizontal line marks the true value
  corresponding to perfect reconstruction. \label{fig:reconstruction}}
\end{figure*}

In Figs.~\ref{fig:reconstruction}a and e are shown the results of this
procedure with the political blogs and openflights networks (see
Appendix~\ref{app:data}). As a baseline, in both figures we show the
direct similarity $S(\x,\A^*)$ of the data obtained with $n=1$ to the
true network $\A^*$, as dashed curves. In both cases the similarity of
the inferred network $S(\hat\A,\A^*)$ to the true network is larger than
the one obtained with the direct observation $S(\x,\A^*)$ for the vast
majority of the parameter range, indicating systematic positive
reconstruction even with single measurements. Expectedly, the quality of
reconstruction increases progressively with a larger number of
measurements $n$, with the similarity eventually approaching
one. Although perfect reconstruction is not possible with single
measurements when the noise is large, it is a noteworthy and nontrivial
fact that the distance to the true network always decreases when
performing it. This is only possible due to the use of a structured
model such as the HDC-SBM that can recognize the structure in the data
and extrapolate from it. If one would use a fully random model in its
place, the similarity would be zero in the entire range, if $n=1$
(although it would improve for $n>1$).

\begin{figure}
  \includegraphics{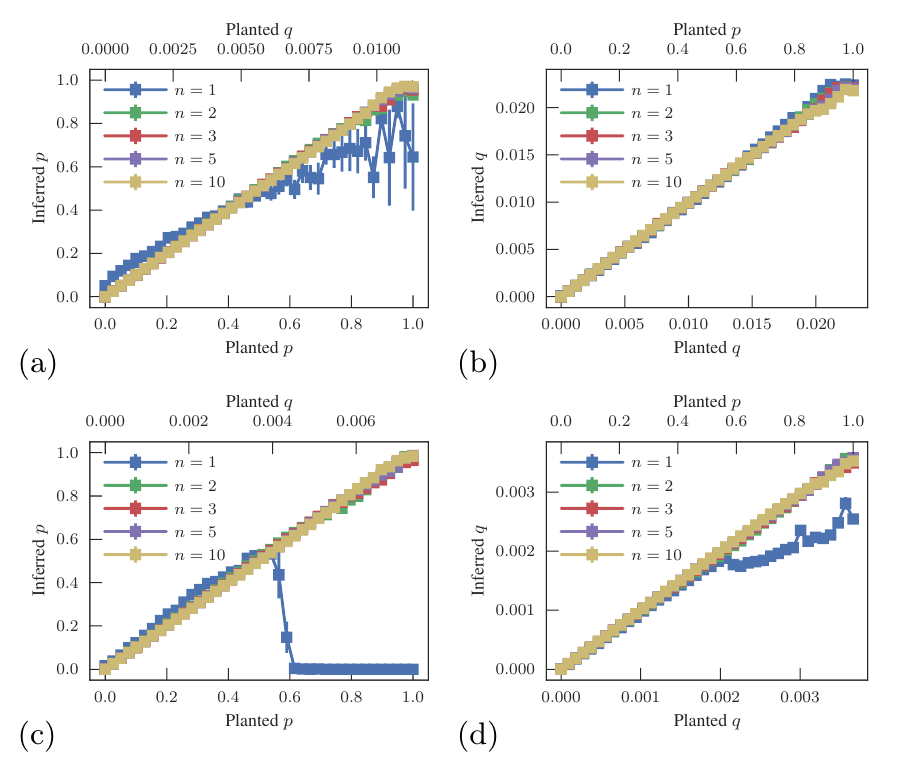}

  \caption{Inferred values of noise magnitude $p$ and $q$ as a function
  of the planted values, for the same simulated measurements described
  in Fig.~\ref{fig:reconstruction}, for the political blogs [(a) and
  (b)] and openflights [(c) and (d)] networks. \label{fig:pq-post}}
\end{figure}

A particularly interesting outcome of the successful reconstruction is
that the noise magnitudes $p$ and $q$ can be determined as well, even
though they are not \emph{a priori} known. As shown in
Fig.~\ref{fig:pq-post} the posterior probability for $p$ and $q$ are
very close to the true values used, even for single measurements. (The
precision of the inferred values of $q$ is generally higher than of $p$,
as we are dealing with sparse networks, with vastly more nonedges than
edges.) For the openflights data the accurate noise recovery only
occurs for moderate magnitudes, and a strong discrepancy is observed for
values around $p\gtrsim 0.5$. In such situations, prior knowledge of the
noise values could have aided the reconstruction for $n=1$, but
otherwise any benefit from this information would have been
marginal. Again, the noise recovery becomes asymptotically exact as we
increase the number of measurements, and is already very accurate for
$n=2$.

We note that the results of Fig.~\ref{fig:reconstruction} remain largely
unchanged if the underlying network considered is sampled from the
DC-SBM with parameters inferred from the original data (not shown).

\subsubsection{Estimating summary quantities}

In addition or instead of the network itself, we may want to estimate a
given scalar observable $y(\A)$ that acts as a summary of some
aspect of the network's structure. In this case, we should seek to
minimize the squared error with respect to the true network $\A^*$,
\begin{equation}
  (\hat y - y(\A^*))^2,
\end{equation}
where $\hat y$ is our estimated value. Like before, without
knowing $\A^*$ the best we can do is minimize the squared error averaged
over the posterior distribution,
\begin{equation}\label{eq:yvar}
  \sigma_{\hat{y}}^2 = \sum_{\A}(\hat y - y(\A))^2P(\A|\n,\x).
\end{equation}
Minimizing $\sigma_{\hat{y}}^2$ with respect to $\hat y$ yields the
posterior mean estimator,
\begin{equation}
  \hat y = \sum_{\A}y(\A)P(\A|\n,\x).
\end{equation}
We can also obtain the uncertainty of this estimator by computing its
variance of Eq.~\ref{eq:yvar}, so that the uncertainty of $\hat y$ is
summarized by its standard deviation, $\sigma_{\hat{y}}$.

It is important to emphasize that in general $\hat y\ne y(\hat\A)$, with
$\hat\A$ being the MMP estimator of Eq.~\ref{eq:mmp}. In other words,
the best estimate for $y(\A^*)$ (i.e. with minimal squared error) is not
the same as the value obtained for the best estimate of $\A^*$
(i.e. with minimal distance).

In Figs.~\ref{fig:reconstruction}b, c, f, and g we see the results of
the same experiment described above, where we attempt to recover the
average local clustering coefficient and the degree assortativity of the
original network. As with the similarity, the inferred values are closer
to the true network's. However, in this case the values for $n=1$ are
substantially closer to the true value for a large range of noise
magnitudes, and is often indistinguishable from it. This means that even
in situations where the posterior distribution of inferred networks
yields a relatively poor similarity to the true network, as it cannot
precisely correct the altered edges and nonedges, it still shares a high
degree of statistical similarity with it, and can accurately reproduce
these summary quantities.

\subsubsection{Estimating degree distributions}

We can also estimate degree distributions $\hat p_k$, defined
as the probability that a node has degree $k$, by treating them like a
collection of scalar measurements, and minimize the squared error
$\sum_k(\hat p_k - p_k(\A))^2$ averaged over the posterior distribution,
which yields the same posterior mean estimator used so far,
\begin{equation}
  \hat p_k = \sum_{\A}p_k(\A) P(\A|\x,\n).
\end{equation}
The same estimator is also obtained when minimizing the Kullback-Leibler (KL)
divergence,
\begin{equation}
 \operatorname{KL}(p(\A)||\hat p) = \sum_kp_k(\A)\ln\frac{p_k(\A)}{\hat p_k},
\end{equation}
over the posterior, which offers a more convenient way to compare
distributions, as it can be interpreted as the amount of information
``lost'' when $\hat{p}_k$ is used to approximate $p_k(\A)$.

For the estimation of the degree probabilities $p_k(\A)$ for each
individual network sampled from the posterior, we model the degrees
$\bm{k}=\{k_i\}$ as a multinomial distribution\footnote{This model is
somewhat crude, as degrees of simple graphs need to be further
constrained~\cite{havel_poznamka_1955, hakimi_realizability_1962}, but
it serves our main purpose of evaluating reconstruction quality.}
\begin{equation}
  P(\bm k |\{p_k\}) = \frac{N!\prod_kp_k^{n_k}}{\prod_kn_k!},
\end{equation}
where $n_k$ is the number of nodes of degree $k$.
The probabilities themselves are modelled by a uniform
Dirichlet mixture, i.e., sampled uniformly from a simplex constrained by
the normalization $\sum_{k=0}^{K}p_k=1$,
\begin{equation}
  P(\{p_k\}) = K!\delta\left(\textstyle\sum_kp_k-1\right),
\end{equation}
where $K$ is the largest possible degree. With this, the the posterior
mean becomes
\begin{equation}
  p_k(\A) = \frac{n_k + 1}{N + K + 1}.
\end{equation}
This estimation is superior to the more naive $p_k=n_k/N$, as it is less
susceptible to statistical fluctuations due to lack of data, such as
when $n_k=0$, although it approaches it for $N\gg K$ and $n_k\gg 1$.

\begin{figure}
  \includegraphics{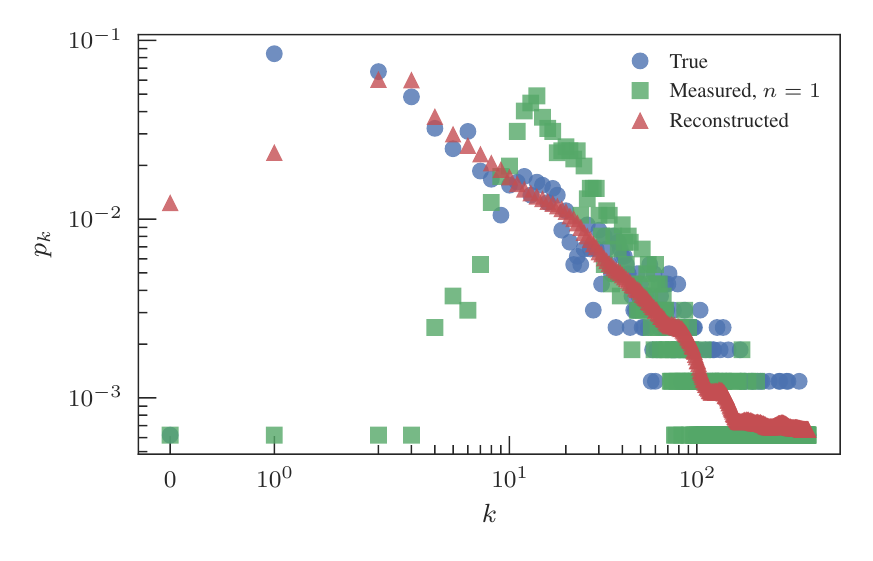}

  \caption{True, measured (with $n=1$) and reconstructed degree
  distributions of the political blog network, with noise magnitudes
  $(p,q)=(0.41, 0.0094)$.\label{fig:deg-dist}}
\end{figure}

In Figs.~\ref{fig:reconstruction}d and h are shown the KL divergence
between the inferred and true distributions, for the same experiments as
before. Like with the local clustering and assortativity coefficients,
the reconstructed degree distributions remain very close to the true
one, despite the continuously decreasing similarity for larger noise
magnitudes. In Fig.~\ref{fig:deg-dist} can be seen the true, measured
and reconstructed distributions for the political blogs network, for a
value of $(p,q)=(0.41, 0.0094)$. Despite the relatively high noise
magnitudes, a single measurement of the network does fairly well in
reconstructing the original distribution, failing mostly only for
degrees zero and one, despite the significant distortion caused by the
noisy measurement process.

\subsubsection{Edge prediction: network de-noising and completion}

\begin{figure}
  \includegraphics{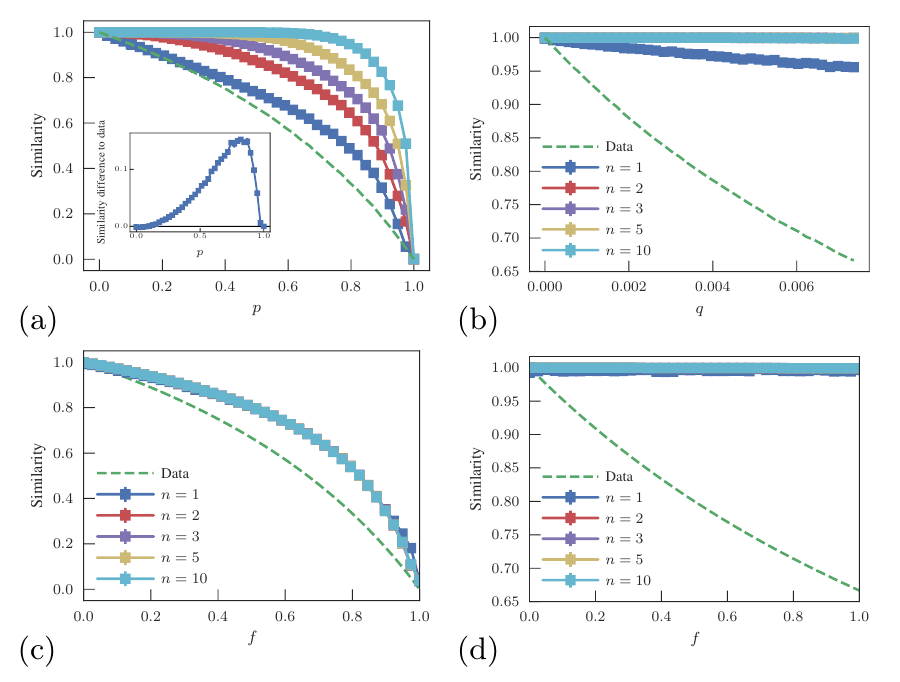}

  \caption{(a) Edge de-noising reconstruction performance for the
  openflights data, as a function of the missing edge probability $p$,
  for various $n$, and $q=0$. The dashed curve shows the corresponding
  value obtained directly with the measured data with $n=1$, and the
  inset shows the difference between the curve for $n=1$ and the dashed
  curve. (b) Same as (a) but for nonedge de-noising, with $p=0$. The
  values of $q$ were chosen to yield the same number of affected
  nonedges as edges in (a). (c) Edge completion reconstruction
  performance as a function of fraction $f$ of unobserved edges. The
  dashed line shows the value of similarity obtained by considering the
  unobserved edges as nonedges. (d) Same as (c) but for nonedge
  completion, as a function of the fraction $f$ of unobserved
  nonedges. The dashed line shows the value of similarity obtained by
  considering the unobserved nonedges as edges.\label{fig:completion}}
\end{figure}

The reconstruction task we have been considering shares many
similarities with the task of model-based edge
prediction~\cite{clauset_hierarchical_2008,guimera_missing_2009}, but is
also different from it in some fundamental aspects. Most typically, edge
prediction is formulated as a binary classification
task~\cite{lu_link_2011}, in which to each missing (or spurious) edge is
attributed a ``score'' (which may or may not be a probability), so that
those that reach a pre-specified discrimination threshold are classified
as true edges (or true nonedges). This threshold is an input of the
procedure, and usually the quality of the classification is assessed by
integrating the true positive rate versus the false positive rate
[a.k.a. the Receiver Operating Characteristic (ROC) curve] for all
discrimination threshold values. This yields the Area Under the Curve
(AUC), which lies in the unit interval, and can be equivalently
interpreted as the probability that a randomly selected true positive
will be ranked above a randomly chosen true negative. Thus, a value of
$1/2$ indicates a performance equivalent to a random guess, and a value
of $1$ indicates ``perfect'' classification (note that a classifier with
AUC value of $1$ still requires the correct discrimination threshold as
an input to fully recover the network).

In contrast, the reconstruction task considered here yields a full
posterior distribution $P(\A|\n,\x)$ for the inferred network
$\A$. Although this can be used to perform the same binary
classification task, by using the posterior marginal probabilities
$\pi_{ij}$ as the aforementioned ``scores,'' it contains substantially
more information. For example, the number of missing and spurious edges
(and hence the inferred probabilities $p$ and $q$) are contained in this
distribution, and thus do not need to be pre-specified. Indeed, our
method lacks any kind of \emph{ad hoc} input, such as a discrimination
threshold (note that the threshold $1/2$ in the MMP estimator of
Eq.~\ref{eq:mmp} is a derived optimum, not an input). This means that
absolute assessments such as the similarity of Eq.~\ref{eq:similarity}
can be computed instead of relative ones such as the AUC.

Furthermore, the reconstruction approach can be used to recover summary
quantities and perform error estimates, which is usually not directly
possible in the binary classifier framing. In addition, reconstructed
networks can contain spurious and missing edges simultaneously, whereas
with traditional edge prediction methods, they each require their own
binary classification (with their own discrimination thresholds).

When doing edge prediction, one often distinguishes recovering from the
effects of noise (i.e. an edge has been transformed into a nonedge, or
vice versa) --- to which we refer as \emph{de-noising} --- and from a
lack of observation (i.e. a given entry in the adjacency matrix is
unknown)
--- to which we refer as \emph{completion}. In each scenario the scores
are computed differently, yielding different classifiers. When
performing reconstruction with our method, we inherently allow for any
arbitrary combination of de-noising and completion: if an entry is not
observed, it has a value of $n_{ij}=0$, which is different from it being
observed with $n_{ij}>0$ as a nonedge $x_{ij}=0$. If the noise
parameters $p$ and $q$ are zero, recovery via the posterior distribution
amounts to a pure completion task for the entries with $n_{ij}=0$, and
likewise we have a pure de-noising task if $n_{ij}>0$ for every pair
$(i,j)$, otherwise we have a mixture of these two tasks.

In Fig.~\ref{fig:completion} we illustrate some of these tasks,
performed using our framework for the openflights dataset, which we
found to be representative of the majority investigated. In
Fig.~\ref{fig:completion}a and b are shown the results for edge ($q=0$)
and nonedge ($p=0$) de-noising, respectively. Given that this network
is sparse, the probability of an edge is on average much smaller than
that of a nonedge, which means that the edge de-noising task is
significantly harder than nonedge de-noising, for which very high
accuracy can be obtained even for $n=1$ measurement per
edge. Nevertheless, positive reconstruction is possible in each case,
approaching a similarity of $1$ as the number of measurements is
increased.

We also perform network completion by choosing a fraction $f$ of edges
or nonedges, for which zero measurements are performed, $n_{ij}=0$,
while the remaining entries are observed $n$ times, $n_{ij}=n$. In
Fig.~\ref{fig:completion}c and d are shown the reconstruction results
for edge and nonedge completion, respectively. Like for de-noising,
nonedge completion is easier, approaching near perfection for the
entire range of parameters, and for the same reason as before. For the
completion tasks, however, the number of observations $n$ for the
non-affected entries has a negligible effect in the reconstruction, and
we observe near-optimal performance already for $n=1$.

Although the number of edges and nonedges affected is the same for both
our de-noising and completion examples, the latter yields a larger rate
of successful reconstruction for both edges and nonedges. This is
understood by noting that these tasks have a different number of
unknowns. In the case of edge completion, on the one hand, for a given
finite fraction $f$ of non-observed edges, we have $O(E)$ unknowns,
which for sparse networks is $O(N)$. For edge de-noising, on the other
hand, for any fraction $p$ of missing edges, for sparse networks we have
in principle $O(N^2)$ possibilities for their placements, corresponding
to all observed nonedges. For nonedge de-noising and completion, the
difficulty is comparable: For any fraction $f = O(1/N)$ left unobserved,
or $q=O(1/N)$ transformed into spurious edges, there are $O(N)$
unknowns, if the network is sparse. However, the actual number of
unknowns for nonedge completion is strictly smaller, as it must involve
only the fraction not observed, whereas for de-noising it involves every
observed edge.

This difference in performance shows how the correct interpretation of
the data can be crucial --- as absence of evidence is not evidence of
absence. Unfortunately, most available datasets fail to make this
distinction, including those few which actually provide some amount of
error assessments, as they do not indicate which pairs of nodes have not
been measured at all.

\subsubsection{Detectability of modular structures}\label{sec:detectability}
\begin{figure}
  \includegraphics{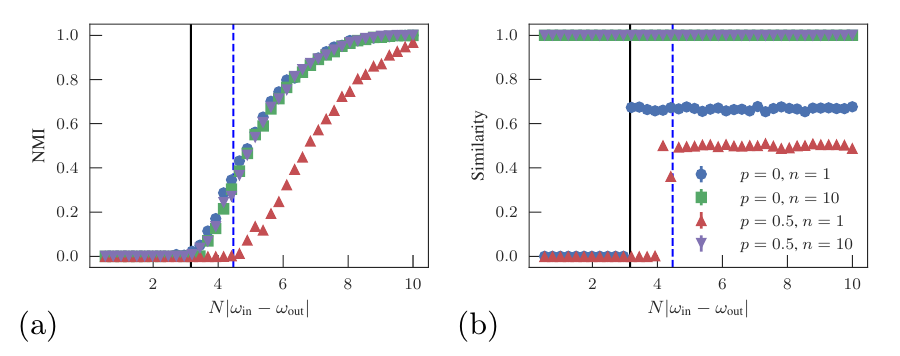}

  \caption{(a) Normalized mutual information (NMI) between
  planted and inferred partitions for a PP model with $N=10^4$, $B=2$,
  $\avg{k}=10$, and measurement errors $q=0$ and $p$ given in the
  legend, together with the number of measurements $n$. The black solid
  line marks the threshold of Eq.~\ref{eq:pp_transition}, and the blue
  dashed line the threshold of Eq.~\ref{eq:pp_transition_noise} with
  $(p,q)=(1/2,0)$. (b) Same as in (a), but for the similarity
  $S(\hat\A,\A^*)$ between the inferred and true
  networks.\label{fig:pp}}
\end{figure}

\begin{figure*}
  \includegraphics{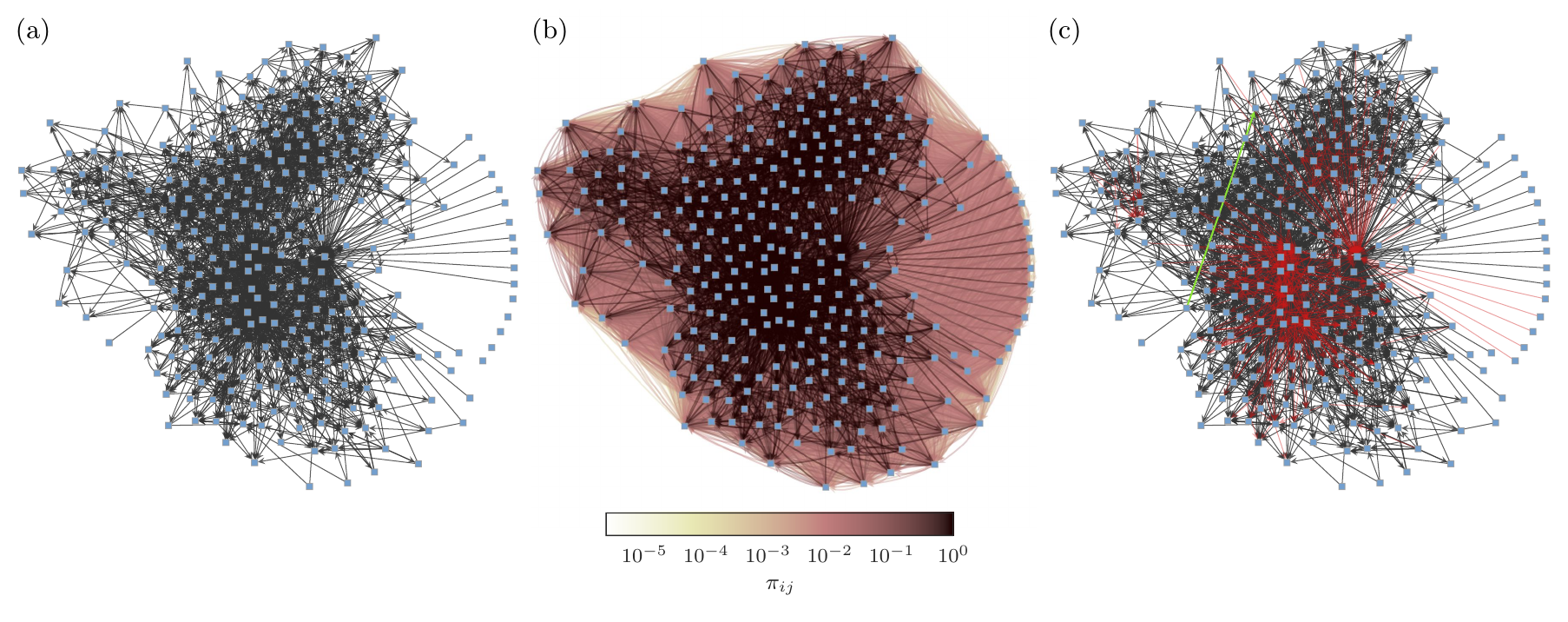}

  \caption{(a) Measured neural network of the \emph{C. elegans}
  worm~\cite{white_structure_1986}. (b) Marginal posterior distribution
  $\pi_{ij}$ of the edges according to our reconstruction method, shown
  as edge colors. (c) Maximum marginal posterior (MMP) estimate of the
  network, with inferred missing edges shown in red, and spurious edges
  shown in green.\label{fig:celegans}}
\end{figure*}

Our approach generalizes the task of community detection for networks
with measurement errors. However, even in the case of error-free
networks with planted community structure, this task is not always
realizable. This is most often illustrated with a simple SBM
parametrization known as the planted partition model (PP),
\begin{equation}
  \omega_{rs} = \omega_{\text{in}}\delta_{rs} + \omega_{\text{out}}(1-\delta_{rs}),
\end{equation}
with equal-sized groups, $n_r=N/B$. As has been shown in Ref.~\cite{decelle_inference_2011},
the detection of communities from networks sampled from this model
undergoes as phase transition, and becomes impossible for parameter
values satisfying
\begin{equation}\label{eq:pp_transition}
  N|\omega_{\text{in}} - \omega_{\text{out}}| < B\sqrt{\avg{k}},
\end{equation}
where $\avg{k} = N[\omega_{\text{in}} + (B-1)\omega_{\text{out}}]/B$ is
the average degree of the network. This transition means that even
though a PP model may contain assortative community structure with
$\omega_{\text{in}} > \omega_{\text{out}}$, the individual samples from
the generative model will be indistinguishable from a fully random graph
if the inequality of Eq.~\ref{eq:pp_transition} is fulfilled, and hence
will contain no information useful for the recovery of the planted
communities.

When considering measured networks, it is expected that the introduced
errors will make the detection task more difficult, as the noise will
remove information from the data. As we have seen in
Sec.~\ref{sec:single}, when a single measurement of a SBM network is
made with noise parameters $p$ and $q$, it becomes indistinguishable
from a SBM sample with effective probabilities $\bm{\omega}'$, given by
Eq.~\ref{eq:sbm_eff}. Applying this to the PP model, yields a
transition according to
\begin{equation}\label{eq:pp_transition_noise}
  N|\omega_{\text{in}} - \omega_{\text{out}}| < \frac{B\sqrt{(1-p-q)\avg{k} + qN}}{(1-p-q)}.
\end{equation}
For positive error magnitudes $p>0$ or $q>0$, the above threshold will
be larger than Eq.~\ref{eq:pp_transition}.  This highlights how
measurement noise can hinder the detection of large-scale structures if
they are sufficiently weak, and induce a phase transition in their
detection. This also means that the reconstruction of the networks
themselves will be affected by the same transition, as our approach
hinges on the detectability of these large-scale structures.

In Fig.~\ref{fig:pp} are shown the reconstruction results for PP network
samples with $B=2$ groups, for simulated measurements always using
$q=0$, but with either $p=0$ or $p=1/2$. Without measurement noise,
$p=0$, the detectability of the planted partition is possible all the
way down to the detectability threshold of
Eq.~\ref{eq:pp_transition}. Despite the lack of noise, the similarity
with the true network is only slightly above $0.6$ in the detectable
region. This is because the probabilities in this ensemble are not
sufficiently heterogeneous to rule out high noise values, as some of the
empirical networks we have considered. Below the transition, the
similarity falls to zero, as the network becomes indistinguishable from
a fully random one. Interestingly, this partial uncertainty about the
network does not affect the inference of the node partition. If we
increase the noise to $p=1/2$, the partition recovery is possible up to
the threshold of Eq.~\ref{eq:pp_transition_noise} when only $n=1$
measurements are made. However, after sufficiently increasing $n$, the
effects of noise are diminished, and the original threshold can be
achieved. In this case, the similarity also becomes high even below the
detectability threshold, where the community structure itself cannot be
recovered. This is because the repeated measurements themselves yield
sufficient information about the network structure, and the
reconstruction no longer needs to rely on the network structure itself.

\subsection{Reconstruction of empirical data and uncertainty assessment}\label{sec:empirical}

A central advantage of our method is that it can be used to reconstruct
noisy networks when only a single measurement has been made for each
entry in the adjacency matrix, and no error assessment is known. As the
majority of network data can be cast into this framework, our method can
be used to reconstruct them and give uncertainty assessments for
quantities of interest. In this section we discuss a few empirical
examples.

We focus first on the neural network of the \emph{Caenorhabditis
elegans} worm. It has been used extensively as a model organism, and it
had its full neural network mapped in 1986 by White et
al~\cite{white_structure_1986}. The network measurement has been done by
electron microscopy of transverse serial sections of the animal's body
of about 50 nm thickness, amounting to around 8000 images. Based on
these images, the network was reconstructed by painstaking manual
tracing of the neuron paths across the different images. The reliability
of the reconstruction procedure was discussed in
Ref.~\cite{white_structure_1986}, where human error in tracing the
neuron bundles, the orientation of the neurons with respect to the
transverse section, and poor image quality were identified as the main
sources of potential errors. White et al. employed a series of error
mitigating procedures, such as detecting basic connection
inconsistencies, exploiting the partial bilateral symmetry for suspect
connections, and comparing with independent reconstructions of parts of
the network. Although the authors of that work profess to be
``reasonably confident'' that the structure they present is
``substantially correct,'' they do not exclude the possibility of
remaining errors, nor do they quantify in any way the uncertainty of
their measurements. Furthermore, the data commonly used for network
analysis, which we also use here, has been manually compiled by Watts et
al.~\cite{watts_collective_1998}, based on the original data of
Ref.~\cite{white_structure_1986}, and may contain further errors. The
resulting data we use amount to $N=302$ nodes and $E=2,345$ directed
edges (note that five nodes were excluded in
Ref.~\cite{watts_collective_1998} for not having any connections. We
include these nodes in our analysis, as it is suspicious that isolated
neurons can exist, and thus is probably a symptom of missing data).

\begin{figure}
  \includegraphics{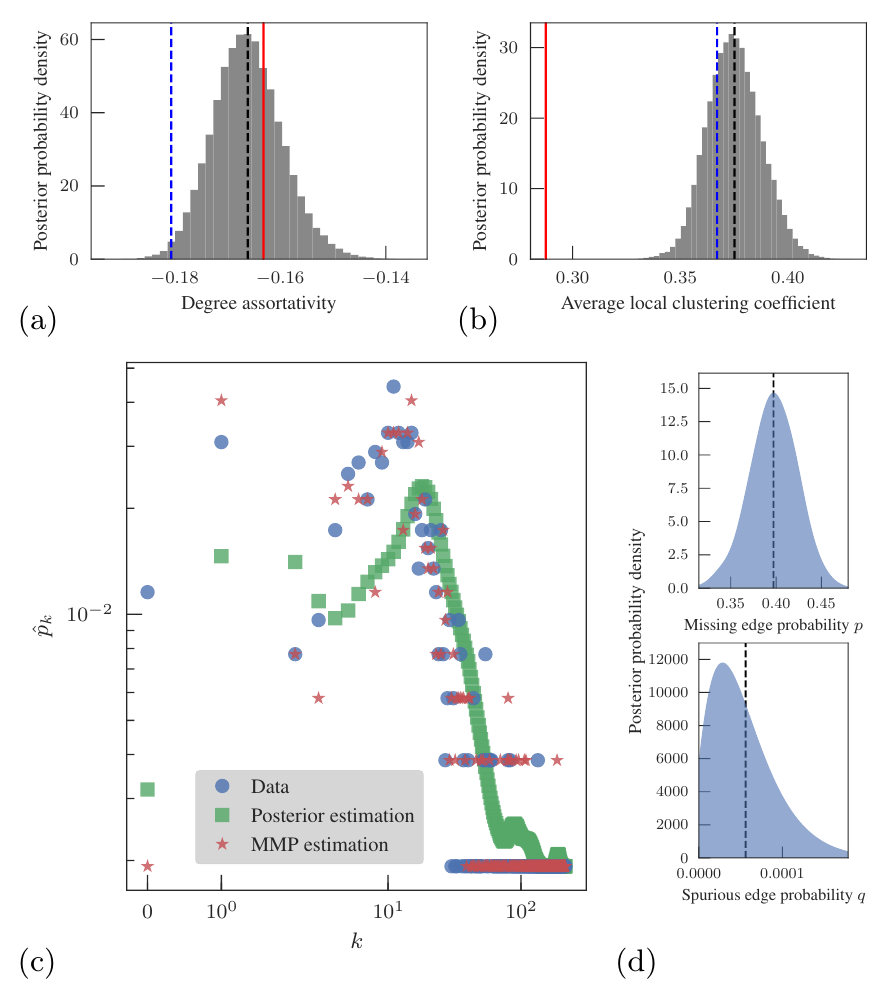}
  \caption{Reconstruction statistics for the neural network of
  \emph{C. elegans}. (a) Posterior distribution of the degree
  assortativity coefficient. The black dashed line marks the mean of the
  distribution, and the blue dashed line the value obtained for the MMP
  estimate, $\hat\A$. The red solid line marks the value computed
  directly from the data. (b) Same as
  (a) but for the average local clustering coefficient. (c) Measured and
  estimated degree distributions. (d) Posterior
  distributions for the error probabilities $p$ and $q$.\label{fig:celegans-dist}}
\end{figure}

When we employ our reconstruction procedure on the \emph{C. elegans}
data, we find the results shown in Figs.~\ref{fig:celegans}
and~\ref{fig:celegans-dist}, and summarized in
Table~\ref{tab:empirical}. The MMP estimate of this network contains
$\hat E = 2,773$ edges, but the posterior distribution is significantly
broad, and contains on average $\avg{E} = 3,950$ edges, meaning that
there are many potential edges with low but non-negligible
probabilities. We note that our reconstruction connects the isolated
nodes in the data to the main hub in the network, which is an important
neuron situated in the head of the worm. As seen in
Fig.~\ref{fig:celegans-dist}a the inferred degree assortativity
coefficient is compatible with the value measured directly from data,
and our method is capable of providing a confidence interval for this
estimation. The same is not true for the average local clustering
coefficient, as seen in Fig.~\ref{fig:celegans-dist}b, which is not
compatible with the value measured directly from data with any
reasonable confidence.

\begin{figure}
  \includegraphics{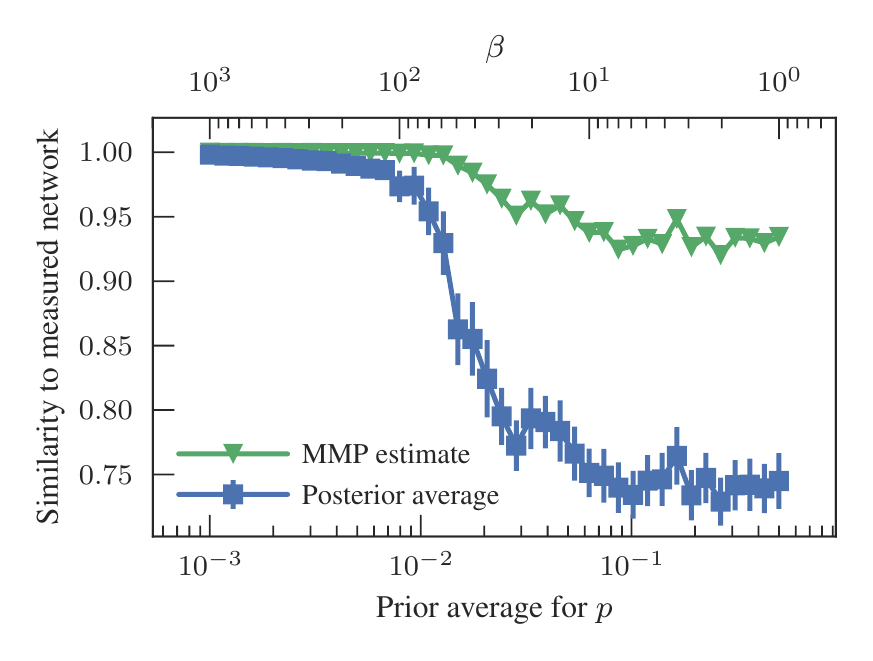}

  \caption{Average similarity between the posterior samples and the measured
      \emph{C. elegans} data as function of the hyperparameter $\beta$
      (with $\alpha=1$), which controls the prior belief on the
      probability $p$ of missing edges (the average of which is shown
      in the x axis). For reference, the similarity for the MMP estimate
      is also shown.\label{fig:celegans_priors}}
\end{figure}

For the \emph{C. elegans} data, the inferred error rates are $(\hat
p,\hat q)=(0.4, 6\times 10^{-5})$. Although this corresponds to a very
high accuracy with respect to spurious edges, it indicates a low
accuracy with respect to missing edges, and it implies that almost half
of the original edges were misrepresented as nonedges. Although the
consensus of the posterior distribution (represented by the MMP
estimate) is reasonably close to the original data, with a similarity of
$0.93$, the similarity averaged over the posterior distribution is only
$0.74$ indicating a fair amount of uncertainty.  This seems to
contradict the qualitative assessment of
Ref.~\cite{white_structure_1986}, which argued in favor of the
reliability of their data. This discrepancy can be interpreted in two
ways: 1. The assessment in Ref.~\cite{white_structure_1986} was too
optimistic, and the data contains indeed more errors than anticipated;
2. The data actually contains fewer errors than our method predicts, but
the true network itself is not sufficiently structured to \emph{rule
out} errors in a manner that can be exploited by our method. However,
even if case 2 happens to be true, our method correctly projects an
agnostic prior assumption about the error rates onto the posterior
distribution, after being informed by the data. This means that more
confidence on the data and the existence of fewer errors must be
accompanied by either more data (e.g. repeated measurements), or a more
refined prior information on the error rates, obtained either by
calibration or a quantitative study of the methods employed in
Ref.~\cite{white_structure_1986}. As an illustration, in
Fig.~\ref{fig:celegans_priors} is shown the posterior similarity with
the date obtained with different choices of the hyperparameter $\beta$,
using $\alpha=1$, which control the prior knowledge on the value of $p$,
with an average given by $\avg{p} = \alpha/(\alpha + \beta)$. A high
accuracy of the data, with inferred similarities approaching one, is
only achieved by a prior belief on $p$ being on the order of $0.01$ or
smaller. This means that one should trust the claimed high accuracy in
Ref.~\cite{white_structure_1986} only if one is confident that the
probability of an edge not being recognized as such was below one
percent. This might very well be true, but would need to be
substantiated with further evidence. Although in situations such as
these our method cannot fully resolve the discrepancy without further
data, it serves as the appropriate framework in which to place the
issue, and shows that any analysis that takes the original measured data
for granted, ignoring potential errors, inherently assumes more
reliability than can be inferred from the data alone.

\begin{figure}
  \includegraphics{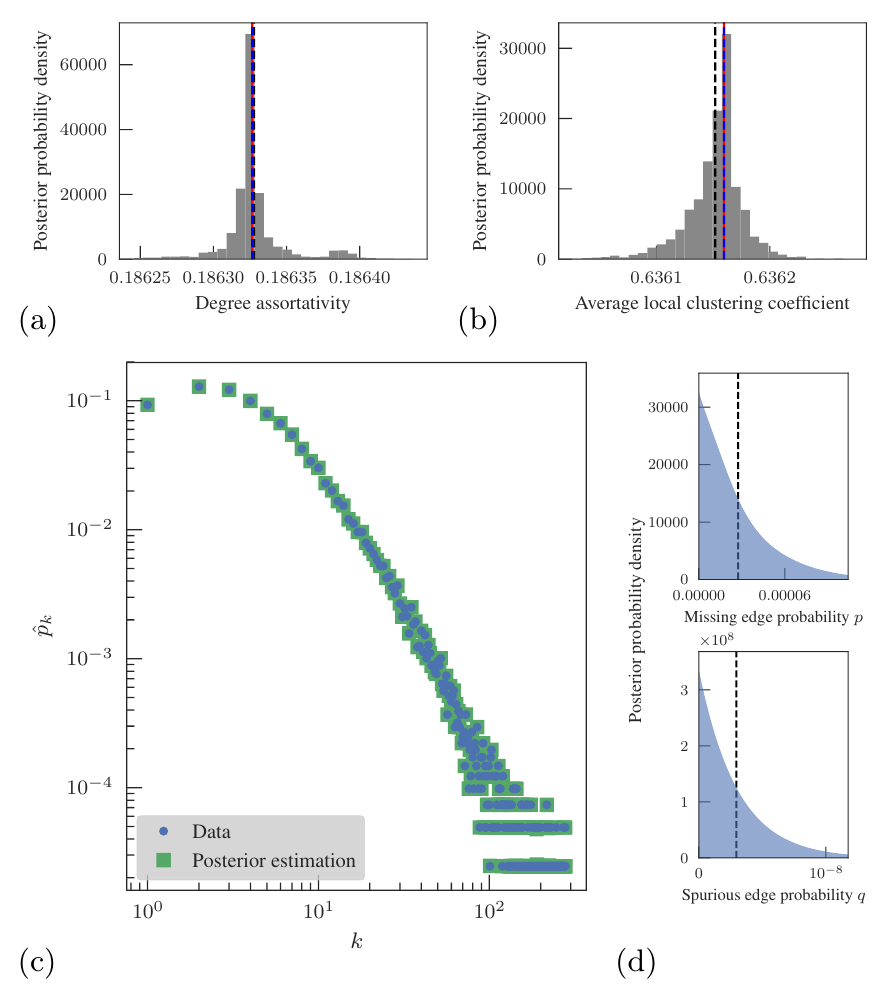}

  \caption{Reconstruction statistics for the co-authorship network of
  \url{arxiv.org}. (a) Posterior distribution of the degree
  assortativity coefficient. The black dashed line marks the mean of the
  distribution, and the blue dashed line the value obtained for the MMP
  estimate, $\hat\A$. The red solid line marks the value computed
  directly from the data. (b) Same as
  (a) but for the average local clustering coefficient. (c) Measured and
  estimated degree distributions. (d) Posterior
  distributions for the error probabilities $p$ and $q$.\label{fig:cond-mat}}
\end{figure}

For other kinds of data, it is possible to obtain very accurate
reconstructions with single measurements. As an example, we consider the
network of collaborations in papers published in the \texttt{cond-mat}
section of the \url{arxiv.org} pre-print website in the period spanning
from January 1, 1995 and March 31, 2005, where authors are nodes, and an
edge exists if two authors published a paper
together~\cite{newman_structure_2001}. This network was compiled by
crawling through the website interface, and could contain errors due to
incorrect parsing.\footnote{These kinds of data also tend suffer from
name ambiguity problems, where the same author appears under different
names, due, for example, to alternative spellings. But since this causes
node duplications to occur, it cannot be corrected with our method,
which can address only spurious and missing edges.} When reconstructed
using our method, however, we find that it is remarkably accurate, with
very low error rates inferred as $(p,q)=(3\times 10^{-5}, 3\times
10^{-9})$. As can be seen in Fig.~\ref{fig:cond-mat}, all inferred
properties match very closely the direct measurement --- although our
reconstruction is still useful in providing error estimates for them.

In Table~\ref{tab:empirical} we provide a summary of reconstruction
results with our method to several empirical networks. We observe a
tendency of larger networks to be more accurate than smaller ones. This
is not a trivial result of there being more data, but rather of these
larger networks containing stronger structures which are informative of
low measurement noise. If these networks were fully random, their
reconstruction accuracy would have been very poor, regardless of their
size.

\begin{table*}
  \resizebox{\textwidth}{!}{
  \smaller
  \setlength{\tabcolsep}{4pt}
  \sisetup{round-mode=places, round-precision = 5}
  \begin{tabular}{l|S[table-format=1.9]|S[table-format=6.0]|S[table-format=7.0]|S[table-format=7.5]|S[table-format=2.5]|S[table-format=2.9]|S[table-format=1.5]|S[table-format=1.8]|S[table-format=4.5]|S[table-format=1.4e2]|S[table-format=1.5e3]}\toprule
    {\multirow{2}{*}{\vspace{-7pt}Dataset}} & {\multirow{2}{*}{\vspace{-7pt}Similarity}} & {\multirow{2}{*}{\vspace{-7pt}Nodes}} & \multicolumn{2}{c|}{Edges} & \multicolumn{2}{c|}{Degree assortativity} & \multicolumn{2}{c|}{Local clustering} & {\multirow{2}{*}{\vspace{-7pt}$B_e$}} & {\multirow{2}{*}{\vspace{-7pt}$\hat p$}}  & {\multirow{2}{*}{\vspace{-7pt}$\hat q$}}\\
    &&&\multicolumn{6}{c}{\vspace{-7pt}}&&&\\ \cline{4-9}
    &&&&\\[-7pt]
    &&&{Direct}&{Estimated}&{Direct}&{Estimated}&{Direct}&{Estimated}&&&\\
\midrule
Karate club& 0.94 +- 0.04 & 34 & 78 & 77 +- 7 & -0.4756130977 & -0.49 +- 0.05 & 0.5706384782 & 0.58 +- 0.05 & 2.7 +- 0.6 & 0.06 +- 0.05 & 0.012 +- 0.01 \\
9/11 terrorists& 0.96 +- 0.02 & 62 & 152 & 154 +- 8 & -0.0804756618 & -0.096 +- 0.02 & 0.4863715559 & 0.5 +- 0.02 & 5.4 +- 0.5 & 0.05 +- 0.04 & 0.003 +- 0.002 \\
American football& 0.857 +- 0.016 & 115 & 613 & 500 +- 18 & 0.1624422496 & 0.18 +- 0.07 & 0.403216011 & 0.68 +- 0.04 & 12.7 +- 0.3 & 0.05 +- 0.03 & 0.0226 +- 0.0019 \\
Network scientists& 0.9981 +- 0.0017 & 379 & 914 & 915 +- 3 & -0.0816778483 & -0.0823 +- 0.0018 & 0.7412306143 & 0.741 +- 0.003 & 29.6 +- 1.4 & 0.004 +- 0.003 & 3.1 +- 1.9 e-5 \\
\emph{C. elegans} neural& 0.744 +- 0.019 & 302 & 2345 & 3950 +- 160 & -0.1631992103 & -0.167 +- 0.007 & 0.2875228459 & 0.378 +- 0.012 & 17 +- 0.3 & 0.41 +- 0.02 & 6 +- 3 e-5 \\
Malaria genes& 0.9981 +- 0.0015 & 1103 & 2965 & 2973 +- 9 & -0.3001327632 & -0.2997 +- 0.002 & 0 & 0(0) & 30.8 +- 0.3 & 0.004 +- 0.003 & 4 +- 3 e-6 \\
Power grid& 0.8 +- 0.07 & 4941 & 6594 & 9900 +- 1300 & 0.0034569877 & 0.043 +- 0.017 & 0.0801036111 & 0.058 +- 0.007 & 15.6 +- 0.7 & 0.33 +- 0.1 & 2.5 +- 1.9 e-7 \\
Political blogs& 0.965 +- 0.005 & 1222 & 16714 & 17860 +- 190 & -0.221328723 & -0.2226 +- 0.0016 & 0.3202546194 & 0.343 +- 0.005 & 16.6 +- 0.3 & 0.066 +- 0.01 & 4.4 +- 1.7 e-5 \\
DBLP citations& 0.64 +- 0.01 & 12590 & 49744 & 106000 +- 2000 & -0.0457245833 & -0.0559 +- 0.0019 & 0.1171835544 & 0.164 +- 0.007 & 86.4 +- 2 & 0.529 +- 0.011 & 9 +- 5 e-9 \\
Openflights& 0.9916 +- 0.0009 & 3286 & 39430 & 40100 +- 70 & -0.0053116897 & -0.0071 +- 0.0011 & 0.4964685588 & 0.507 +- 0.002 & 117.1 +- 0.5 & 0.0167 +- 0.0018 & 1 +- 0.3 e-7 \\
Reactome& 0.999977 +- 0.00001 & 6327 & 146160 & 146164 +- 3 & 0.2448744809 & 0.24487 +- 0.00004 & 0.588375016 & 0.5887 +- 0.0003 & 318.7 +- 1 & 4.1 +- 1.8 e-5 & 1.3 +- 0.8 e-7 \\
cond-mat& 0.999986 +- 0.000013 & 40421 & 175693 & 175695 +- 4 & 0.1863265789 & 0.18633 +- 0.00002 & 0.6361596433 & 0.63615 +- 0.00003 & 1014 +- 6 & 3 +- 2 e-5 & 3 +- 2 e-9 \\
Enron email& 0.99986 +- 0.00005 & 36692 & 183831 & 183885 +- 18 & -0.1107640326 & -0.11075 +- 0.00002 & 0.4969825596 & 0.49692 +- 0.00008 & 188.9 +- 1.1 & 0.00028 +- 0.0001 & 2.9 +- 1.9 e-9 \\
Linux source& 0.9973 +- 0.0003 & 30837 & 213424 & 214600 +- 120 & -0.1746789741 & -0.17467 +- 0.00007 & 0.128491999 & 0.1322 +- 0.001 & 351.2 +- 0.7 & 0.0055 +- 0.0005 & 1.7 +- 1 e-9 \\
Brightkite& 0.9985 +- 0.0003 & 58228 & 214078 & 214740 +- 80 & 0.0108157963 & 0.011 +- 0.00011 & 0.1723259274 & 0.17234 +- 0.0001 & 151 +- 3 & 0.0029 +- 0.0005 & 1.7 +- 1.2 e-8 \\
PGP& 0.99799 +- 0.00009 & 39796 & 301498 & 301660 +- 60 & 0.0007574928 & 0.00049 +- 0.00008 & 0.4610872993 & 0.4617 +- 0.0002 & 929 +- 2 & 0.00227 +- 0.00016 & 3.35 +- 0.18 e-7 \\
Internet AS& 0.99967 +- 0.00013 & 53387 & 496731 & 497070 +- 130 & -0.1869674555 & -0.186959 +- 0.000017 & 0.6809685281 & 0.68126 +- 0.00014 & 218 +- 1.6 & 0.0007 +- 0.0003 & 1 +- 0.8 e-9 \\
Web Stanford& 0.9999987 +- 0.0000008 & 281903 & 2312497 & 2312494 +- 4 & -0.1124445174 & -0.1124447 +- 0.0000002 & 0.5976304608 & 0.597634 +- 0.000003 & 4168 +- 2 & 1 +- 0.2 e-6 & 7 +- 5 e-11 \\
Flickr& 0.999976 +- 0.000013 & 105938 & 2316948 & 2316830 +- 60 & 0.2468511883 & 0.246823 +- 0.000016 & 0.0891280227 & 0.089138 +- 0.000007 & 617 +- 2 & 6 +- 3 e-7 & 2 +- 1.1 e-8 \\
\bottomrule
  \end{tabular}} \caption{Reconstruction results for empirical networks
  with single measurements per edge, and no available primary error
  assessments. Similarity refers to the average of $S(\A,\A^*)$ over the
  posterior distribution. For each quantity (number of edges, degree
  assortativity, average local clustering) is shown the value directly
  obtained from the data (direct) and the average over the posterior
  distribution (estimated). The value of $B_e$ is the posterior average
  of the effective number of inferred communities $e^{H(\bm{n})}$, with
  $H(\bm{n})=-\sum_r(n_r/N)\ln (n_r/N)$, where $n_r$ is the number of
  nodes in group $r$, being the entropy of the group size
  distribution. The values $\hat{p}$ and $\hat{q}$ are the posterior
  averages of the error rates. In all cases, the parentheses indicate the
  standard deviation over the posterior distribution. Dataset
  descriptions are given in
  Appendix~\ref{app:data}. \label{tab:empirical}}
\end{table*}

\subsection{Heterogeneous errors}\label{sec:heterogeneous}
\begin{figure*}
  \includegraphics{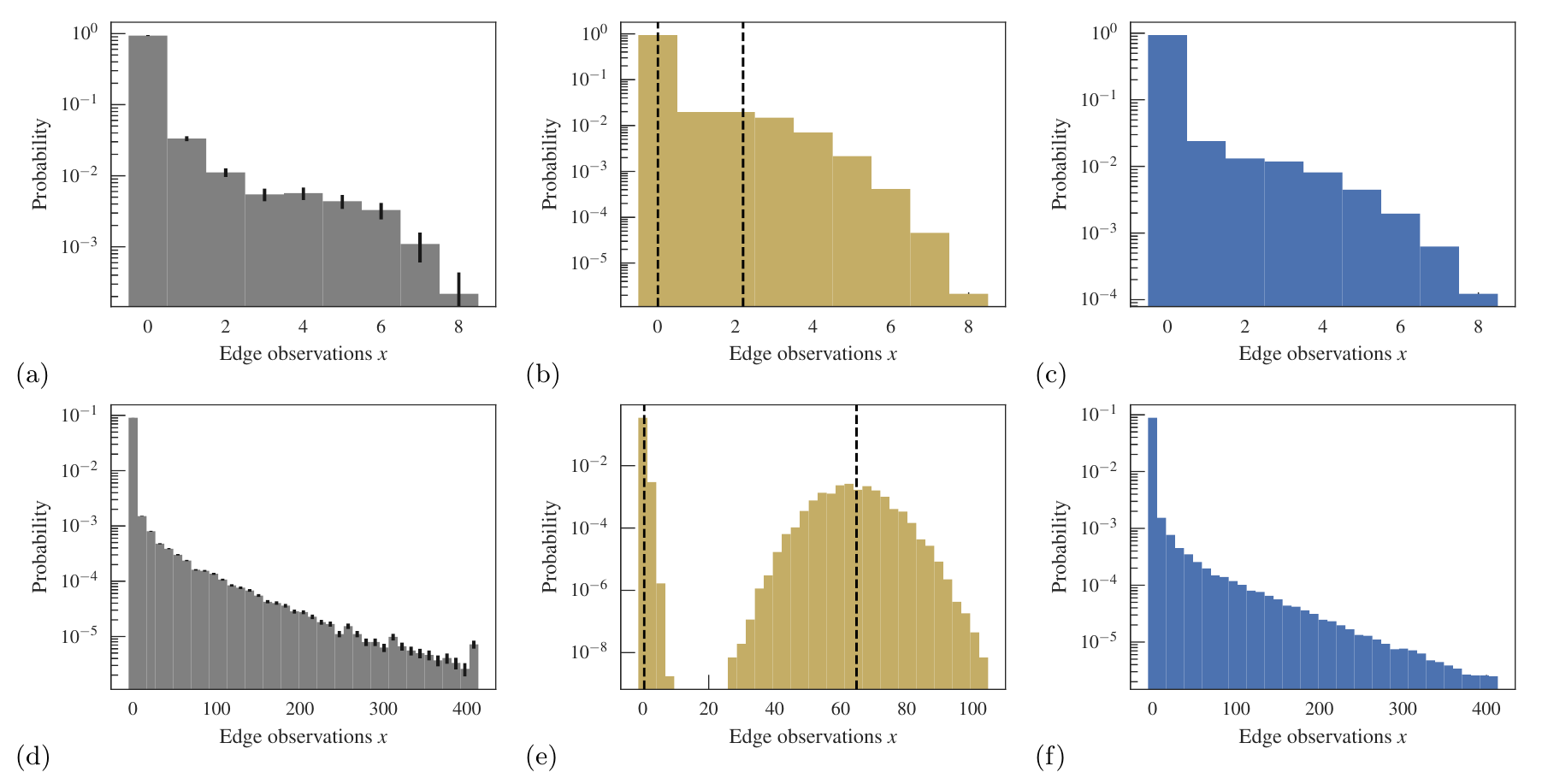}

  \caption{Distribution of edge occurrences, $x_{ij}$, for
  the reality mining (top row) and human connectome (bottom row)
  datasets. [(a) and (d)] Empirical data. [(b) and (e)] Generated from
  inferred parameters, according to the uniform model. [(c) and (f)]
  Generated from inferred parameters, according to the nonuniform
  model.\label{fig:multiple}}
\end{figure*}

\begin{figure*}
  \includegraphics{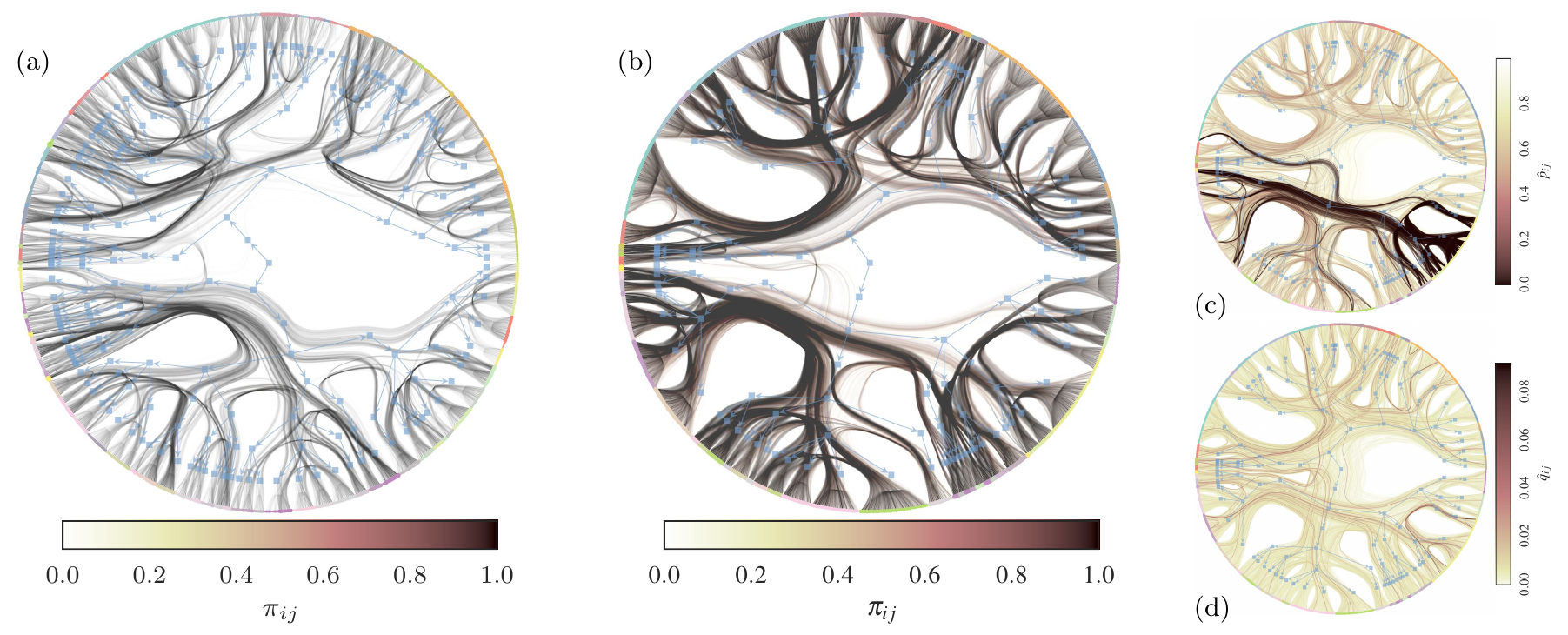}

  \caption{Reconstruction results for the human
  connectome. (a) Marginal posterior distribution of edges $\pi_{ij}$
  and inferred hierarchical partition, according to the model with
  uniform errors. The upper hierarchy branch corresponds to the right
  hemisphere. (b) Same as (a) but with the nonuniform model. (c)
  Inferred missing edge probabilities $p_{ij}$ for the nonuniform
  model. (d) Same as (c) but for the spurious edge probabilities,
  $q_{ij}$.\label{fig:connectome}}
\end{figure*}

So far we have considered only the situation where the error
probabilities $p$ and $q$ are the same for every pair of nodes in the
network. Although it is easy to imagine a simplified scenario where the
same measurement instrument is used in every case, it is also easy to
imagine situations where this is not an adequate representation of how
measurement is made. For example, in the case of the \emph{C. elegans}
neural network, the spatial proximity of the neurons may make it harder
or easier to measure the edges and nonedges, thus impacting their error
probabilities.

With this in mind, it is easy to generalize our framework to allow for
individual error probabilities $p_{ij}$ and $q_{ij}$, for missing and
spurious edges between nodes $i$ and $j$, respectively. Given a true
underlying entry $A_{ij}$ between these two nodes, its measurement
probability is given by
\begin{multline}
  P(x_{ij}|n_{ij}, A_{ij}, p_{ij}, q_{ij}) =\\ {n_{ij}\choose
  x_{ij}}\left[(1-p_{ij})^{x_{ij}}p_{ij}^{n_{ij}-x_{ij}}\right]^{A_{ij}}\times\\\left[q_{ij}^{x_{ij}}(1-q_{ij})^{n_{ij}-x_{ij}}\right]^{1-A_{ij}}.
\end{multline}
Using the same Beta priors as before, we can integrate over $p_{ij}$ and
$q_{ij}$, obtaining
\begin{multline}
  P(x_{ij}|n_{ij}, A_{ij}, \alpha,\beta,\mu,\nu)\\
  =\int P(x_{ij}|n_{ij}, A_{ij}, p_{ij}, q_{ij})P(p_{ij}|\alpha,\beta)P(q_{ij}|\mu,\nu)\;\dd p_{ij}\dd q_{ij}\\
  = {n_{ij}\choose x_{ij}}\left[\frac{\mathcal{B}(n_{ij}-x_{ij}+\alpha, x_{ij}+\beta)}{\mathcal{B}(\alpha,\beta)}\right]^{A_{ij}}\times\\
  \left[\frac{\mathcal{B}(x_{ij}+\mu, n_{ij}-x_{ij}+\nu)}{\mathcal{B}(\mu,\nu)}\right]^{1-A_{ij}}.
\end{multline}
With this we have the full conditional distribution for the measured network,
\begin{equation}
  P(\x|\n,\A,\alpha,\beta,\mu,\nu)=\prod_{i<j}P(x_{ij}|n_{ij}, A_{ij}, \alpha,\beta,\mu,\nu)
\end{equation}
with which we can obtain the posterior distribution of
Eq.~\ref{eq:post_general}. However, unlike the case with uniform errors,
the choice of hyperparameters is now vital. The noninformative
assumption $\alpha=\beta=\mu=\nu=1$ applied above makes the likelihood
\emph{independent} of the planted network $\A$, rendering the data
completely uninformative as well. This means we must have some global
information that specifies how the values of $p_{ij}$ and $q_{ij}$ are
distributed. Although we could simply set (or fit) the values of the
hyperparameters to values different from one, we favor a
nonparametric approach, and we include the hyperparameters in the
posterior distribution,
\begin{multline}
  P(\A, \bb, \alpha,\beta,\mu,\nu|\n,\x) = \\
  \frac{P(\x|\n,\A,\alpha,\beta,\mu,\nu)P(\A|\bb)P(\bb)P(\alpha,\beta,\mu,\nu)}{P(\x|\n)}
\end{multline}
which requires their own hyperprior distribution
$P(\alpha,\beta,\mu,\nu)$. Here we will be agnostic and use a constant
prior $P(\alpha,\beta,\mu,\nu) \propto 1$, with an unspecified and
unnecessary normalization constant, as it cancels out in the posterior
distribution.\footnote{In fact, since $\alpha,\beta,\mu$ and $\nu$ are
unbounded continuous variables, the constant prior cannot be normalized,
making it improper. The way around this is to use instead a constant
prior constrained to some domain of interest, outside of which it is
zero. If this domain is large enough to contain the inferred values, the
resulting posterior will be very close to the one obtained with the
improper prior, which is identical to the limit (if it exists) of the
posterior distribution where the domain boundaries go to
infinity.\label{foot:improper}} The inference algorithm is the same as
before, but in addition to move proposals for the network $\A$ and node
partition $\bb$, we make also move proposals for the hyperparameters.

Like in the uniform case, we can obtain the posterior distribution for
the error probabilities via their conditional posteriors, i.e.
\begin{multline}
  P(p_{ij}|n_{ij},x_{ij},A_{ij},\alpha,\beta)=\\
  \frac{p^{A_{ij}(n_{ij}-x_{ij})+\alpha-1}(1-p)^{x_{ij}A_{ij}+\beta-1}}{\mathcal{B}(A_{ij}(n_{ij}-x_{ij})+\alpha, x_{ij}A_{ij}+\beta)}
\end{multline}
and likewise for $q_{ij}$ with
\begin{multline}
  P(q_{ij}|n_{ij},x_{ij},A_{ij},\mu,\nu)=\\
  \frac{q^{(1-A_{ij})x_{ij}+\mu-1}(1-q)^{(1-A_{ij})(n_{ij}-x_{ij})+\nu-1}}{\mathcal{B}((1-A_{ij})x_{ij}+\mu, (1-A_{ij})(n_{ij}-x_{ij})+\nu)},
\end{multline}
averaged over the posterior distribution.

We note that for heterogeneous error rates, the case with single
measurements $n_{ij}=1$ become less interesting. If we replace $n_{ij}=1$ and
$x_{ij}\in\{0,1\}$ in the above equations, they become identical to
Eq.~\ref{eq:mlikelihood} for the case with uniform errors, if we make
the substitution
\begin{align}
  p &= \frac{\mathcal{B}(\alpha+1,\beta)}{\mathcal{B}(\alpha,\beta)} = \frac{\alpha}{\alpha+\beta},\\
  q &= \frac{\mathcal{B}(\mu+1,\nu)}{\mathcal{B}(\mu,\nu)} = \frac{\mu}{\mu+\nu}.
\end{align}
In this situation, only the prior averages of $p_{ij}$ and $q_{ij}$
matter, not their variance. A uniform prior for $\alpha, \beta, \mu$ and
$\nu$ is equivalent to Beta priors with parameters $(1, 0)$ for $p$ and
$q$ computed via the equation above,\footnote{Note that Beta
distributions with parameters $(1,0)$ are also improper, but will yield
meaningful results for the same reason given in
footnote~\ref{foot:improper}.} and hence this approach becomes
completely identical to the one with uniform errors considered
before. Therefore, there is no sufficient data in the single measurement
case to detect heterogeneous errors of this kind, and thus a meaningful
use of this method is confined to data with $n_{ij} > 1$. Note also that
this implies that any error heterogeneity present in the data will be
conflated with underlying network structure when single measurements are
made. Ultimately, this conflation can only be resolved by making
multiple measurements.

\begin{figure}
  \includegraphics{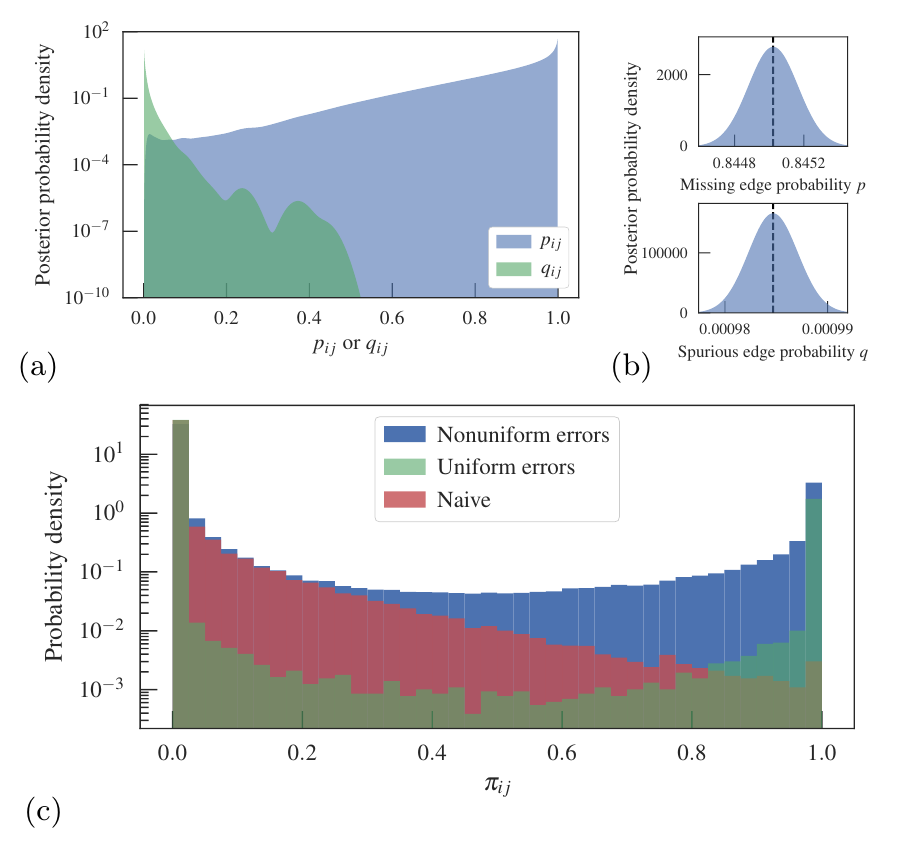}

  \caption{Inferred uncertainties for the human
  connectome. (a) Posterior distribution of $p_{ij}$ and $q_{ij}$, using
  the nonuniform model. (b) Posterior distribution of $p$ and $q$, using
  the uniform model. (c) Distribution of posterior marginal edge
  probabilities $\pi_{ij}$, according to both model variants, as well as
  the naive estimate $\tilde{\pi}_{ij} =
  x_{ij}/n_{ij}$. \label{eq:multiple_posterior}}
\end{figure}

\begin{table*}
  \resizebox{\textwidth}{!}{
  \smaller
  \setlength{\tabcolsep}{4pt}
  \sisetup{round-mode=places, round-precision = 5}
  \begin{tabular}{l|S[table-format=3.0]|S[table-format=4.0]|S[table-format=5.3]|S[table-format=5.5]|S[table-format=2.6]|S[table-format=2.5]|S[table-format=1.7]|S[table-format=1.5]|S[table-format=2.3]|S[table-format=2.3]|S[table-format=1.7]|S[table-format=1.4]|S[table-format=1.5e2]|S[table-format=1.3e1]}\toprule
    {\multirow{2}{*}{\vspace{-7pt}Dataset}} & {\multirow{2}{*}{\vspace{-7pt}$n$}} & {\multirow{2}{*}{\vspace{-7pt}Nodes}} & \multicolumn{2}{c|}{Edges} & \multicolumn{2}{c|}{Degree assortativity} & \multicolumn{2}{c|}{Local clustering} &  \multicolumn{2}{c|}{$B_e$} & \multicolumn{2}{c|}{$\hat{p}$}  & \multicolumn{2}{c}{$\hat{q}$}\\
    &&&\multicolumn{12}{c}{\vspace{-7pt}}\\ \cline{4-15}
    &&&&\\[-7pt]
    &&&{Uniform}&{Nonuniform}&{Uniform}&{Nonuniform}&{Uniform}&{Nonuniform}&{Uniform}&{Nonuniform}&{Uniform}&{Nonuniform}&{Uniform}&{Nonuniform}\\
    \midrule
Karate club& 2 & 34 & 77.9 +- 0.3 & 95 +- 6 & -0.475 +- 0.003 & -0.43 +- 0.05 & 0.569 +- 0.008 & 0.63 +- 0.05 & 2.9 +- 0.6 & 2.9 +- 0.6 & 0.012 +- 0.002 & 0.49 +- 0.03 & 0.0011 +- 0.0003 & 0.0004 +- 0.0013 \\
Reality mining& 8 & 96 & 293 +- 11 & 280 +- 20 & -0.23 +- 0.03 & -0.23 +- 0.03 & 0.31 +- 0.02 & 0.29 +- 0.02 & 3.5 +- 0.6 & 3.4 +- 0.6 & 0.724 +- 0.008 & 0.71 +- 0.03 & 0.0007 +- 0.0002 & 0.001 +- 0.002 \\
School friends& 6 & 2539 & 12500 +- 40 & 8200 +- 300 & 0.258 +- 0.004 & 0.322 +- 0.006 & 0.1535 +- 0.0013 & 0.188 +- 0.003 & 82.5 +- 0.3 & 80.2 +- 0.3 & 0.5064 +- 0.0011 & 0.16 +- 0.03 & 1.8 +- 0.07 e-5 & 0.0002 +- 0.0009 \\
Human connectome& 418 & 1015 & 23020 +- 16 & 62000 +- 6000 & 0.0008 +- 0.0005 & 0.002 +- 0.003 & 0.6796 +- 0.0004 & 0.68 +- 0.07 & 100.5 +- 1.1 & 51.26 +- 0.19 & 0.84503 +- 0.00008 & 0.93 +- 0.09 & 0.0009846 +- 0.000001 & 1 +- 11 e-4 \\
\bottomrule
  \end{tabular}}

  \caption{Reconstruction results for empirical networks with multiple
  measurements per edge. For each quantity is show the value obtained
  using either the uniform or the nonuniform model, as indicated.  The
  value of $B_e=e^{H(\bm{n})}$ is the effective number of inferred
  communities, computed as $H(\bm{n})=-\sum_r(n_r/N)\ln (n_r/N)$, where
  $n_r$ is the number of nodes in group $r$. The values $\hat{p}$ and
  $\hat{q}$ are the posterior averages of the error rates. In all cases,
  the parentheses indicate the standard deviation over the posterior
  distribution.  Dataset descriptions are given in
  Appendix~\ref{app:data}.\label{tab:multiple}}
\end{table*}

We consider two datasets which contain multiple measurements, in order
to compare both approaches. We consider the reality mining dataset,
which recorded proximity interactions between voluntary students over
time~\cite{eagle_reality_2006}. Following
Ref.~\cite{newman_network_2018}, as measurements we considered the state
of the network during eight consecutive Wednesdays in March and April of
2005, so chosen to avoid weekly periodic events. In addition, we
consider the human connectome, using data from the Budapest Reference
Connectome~\cite{szalkai_parameterizable_2017} (which itself is based on
primary data from the Human Connectome
Project~\cite{mcnab_human_2013}). This dataset contain records of the
neuronal connections of $418$ individuals, each of which we considered
as a separate measurement.

For both datasets considered --- as it is arguably always true whenever
multiple network measurements are made --- it is debatable whether there
is really a true single network behind the measurements, as our method
assumes. For example, in the reality mining dataset, the underlying
network could be changing over time, and the connectome can vary between
individuals for physiological reasons, rather than measurement error. In
each case, however, we are free to keep the mathematical structure of
our model in place, and change its interpretation. We could, for
instance, assume that the single network being inferred amounts simply
to a consensus or a blue print of the network, and the ``error'' rates
$p_{ij}$ and $q_{ij}$ indicate the variability of each single edge or
nonedge around this blue print. Since both scenarios are generally
conflated when making this kind of measurement, we can choose the
interpretation that is most suitable according to the context.

In Fig.~\ref{fig:multiple}a and d are shown the distributions of the
measured frequencies of edge occurrences, $x_{ij}$, for both
datasets. For the human connectome, we observe a very broad
distribution, with occurrences present in the entire possible range. In
Fig.~\ref{fig:multiple}b and e we see the simulated results by sampling
parameters from the posterior distribution and generating new data from
them, using in this case the model with uniform errors. Whereas the
results for reality mining are reasonably close to the data, the results
for the human connectome show an obvious discrepancy, where the
generated data is concentrated around two modes, corresponding to the
frequencies of edges and nonedges. Indeed, for the uniform model this
separation is guaranteed to occur for any given $p\ne 1/2$ and $q \ne
1/2$ and a sufficiently large number of measurements. The fact that this
is not observed in the data is a clear indication that the error rates
are not uniform (or alternatively, but mathematically equivalently,
that there is no single network behind the measurements). Indeed when
using the nonuniform model, it recovers the observed frequency almost
perfectly, as seen in Figs.~\ref{fig:multiple}c and f.

If we look more closely at the human connectome data, we see that both
approaches give us different pictures of the underlying network
structure. As is summarized in Table~\ref{tab:multiple}, the uniform
model yields a sparser network, which nevertheless seems more finely
structured, with close to $100$ effective groups detected. Conversely,
the nonuniform model yields a denser network, with a more uniform
structure, and only half as many identified groups. In
Fig.~\ref{fig:connectome} we see more clearly the differences between
both results. Both are capable of uncovering the hemispherical divisions
and the partial bilateral symmetry of the connectome. The nonuniform
model can detect a larger number of edges, but it yields larger
probabilities of missing edges $p_{ij}$ which are heterogeneously
distributed. In Fig.~\ref{fig:connectome}c it can be seen that the
inferred $p_{ij}$ are strongly correlated with the detected group
structure, and in particular seem to indicate a rather stable set of
edges (low $p_{ij}$) that belong mostly to the left hemisphere. The
uniform model, on the other hand, incorporates the variability of edge
occurrences in the model itself, subdividing the groups further to
accommodate it. Therefore, the nonuniform model gives a more faithful
separation between the consensus and the variability around it.

In Fig.~\ref{eq:multiple_posterior} we can see the posterior
distributions of $p_{ij}$, $q_{ij}$ for the nonuniform model, as well
$p$ and $q$ for the uniform model, showing how the former is indeed
significantly more heterogeneous than the latter. In
Fig.~\ref{eq:multiple_posterior}c is also shown the distribution of
posterior probabilities $\pi_{ij}$ for both models, in addition to the
naive estimate $\tilde{\pi}_{ij} = x_{ij}/n_{ij}$. This naive estimate
is crude, as it does not differentiate between the different sources of
error (spurious or missing edge), and does not take into account the
observed correlations between the different entries. Indeed, as the
Fig.~\ref{eq:multiple_posterior}c shows, it leads to very different
results, which are not correctly justified, and should be avoided.

\section{Incorporating extrinsic uncertainty estimates}\label{sec:foreign}

So far we have considered only situations where direct error estimates
on the edges originate from repeated measurements. However, there are
situations where primary error estimates are made under different
formats. Here we consider the scenario of
Ref.~\cite{martin_structural_2016}, where an arbitrary measurement
process is made which yields uncertainty assessments for each node pair,
$Q_{ij} \in [0, 1]$, interpreted as conditionally independent
probabilities, i.e.
\begin{equation}\label{eq:ext_Q}
  P_Q(\A | \Q) = \prod_{i<j}Q_{ij}^{A_{ij}}(1-Q_{ij})^{1-A_{ij}}.
\end{equation}
In principle, we could use these probabilities as they are, and generate
networks and measure their properties from this distribution. But we
could also extract from this information the measurement process which
it represents, and couple it with our reconstruction approach. This
gives us the advantage of being able to use the large scale structure in
the data to better inform our estimates of the underlying network.

\begin{figure}
  \includegraphics{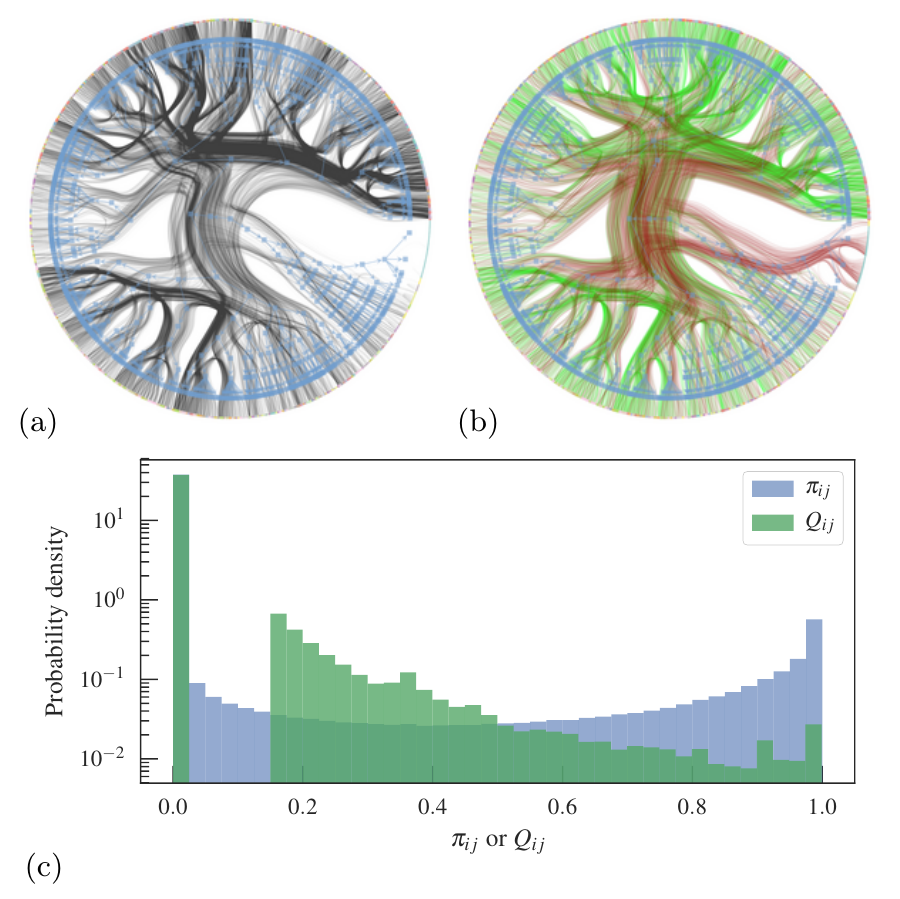}

  \caption{(a) Inferred \emph{E. coli} protein interaction network,
  according to uncertain data $\Q$, using the MMP estimator from the
  posterior $P(\A|\Q)$. (b) Difference between (a) and the MMP estimator
  using the original uncertainties $\Q$ directly, via $P_Q(\A|\Q)$
  (Eq.~\ref{eq:ext_Q}). Green edges are those that are added in (a), and
  red ones are removed. (The hierarchical partition is the same as in
  (a), and is shown only as a visual aid.) (c) Distribution of marginal
  posterior probabilities $\pi_{ij}$ and original uncertainties
  $Q_{ij}$. \label{fig:proteins}}
\end{figure}

The distribution $P_Q(\A | \Q)$ implies the following noisy measurement
process,
\begin{equation}
  P(\Q | \A) = \frac{P_Q(\A | \Q)P_Q(\Q)}{P_Q(\A)},
\end{equation}
with normalization constant
\begin{equation}
  P_Q(\A) = \int P_Q(\A | \Q)P_Q(\Q)\;\dd\Q.
\end{equation}
If we assume the prior on the edge uncertainties are identically
distributed and conditionally independent, i.e.
\begin{equation}
  P_Q(\Q) = \prod_{i<j}P(Q_{ij}),
\end{equation}
we have
\begin{equation}
  P_Q(\A) = \prod_{i<j}\bar{Q}^{A_{ij}}(1-\bar{Q})^{1-A_{ij}},
\end{equation}
with $\bar{Q} = \int_0^1QP(Q)\dd Q$. Combining these together we have
\begin{equation}
  P(\Q | \A) = P_Q(\Q) \prod_{i<j}\left(\frac{Q_{ij}}{\bar{Q}}\right)^{A_{ij}}\left(\frac{1-Q_{ij}}{1-\bar{Q}}\right)^{1-A_{ij}}.
\end{equation}
The above depends on an unknown prior $P_Q(\Q)$. Determining it would
require us to delve into the details of how this measurement is made,
which is unavailable to us if all we know is $P_Q(\A | \Q)$. However
since it is only a multiplicative constant that does not depend on the
data or any latent variable, it will not affect the posterior
distribution, and thus we do not need to determine it. The single aspect
of this distribution that is relevant is its average, $\bar{Q}$. By
allowing only for a minor violation of the Bayesian ansatz, we can
estimate this directly from data
\begin{equation}\label{eq:bar_Q}
  \bar{Q} = \frac{\sum_{i<j}Q_{ij}}{{N\choose 2}}.
\end{equation}
With this, we can couple this arbitrary noise generating process with
our overall framework by taking $\D=\Q$, and obtaining the posterior
distribution
\begin{equation}\label{eq:posterior_Q}
  P(\A|\Q) = \frac{P(\Q | \A)P(\A)}{P(\Q)}
\end{equation}
where $P(\A)$ assumes that the network has been generated by a SBM. Note
that $P(\A|\Q) \ne P_Q(\A|\Q)$, as we are keeping the same noise
generating process, but changing our prior assumption about the data. As
desired, our prior is structured, and is capable of detecting
large-scale patterns --- latent groups of nodes and their probabilities
of connections, as well as node degrees and hierarchical structure ---
to inform our inference. This also highlights the versatility of our
framework, as we are free to replace the measurement model as appropriate.

Although our derivation is somewhat different, equations
Eq.~\ref{eq:ext_Q} to~\ref{eq:bar_Q} above are the same as in
Ref.~\cite{martin_structural_2016}. The resulting posterior of
Eq.~\ref{eq:posterior_Q}, however, is different, as our approach is
nonparametric, and hence can be used to infer the number of groups, and
does not involve any approximations that rely on the network being
sparse or locally tree-like.

In Fig.~\ref{fig:proteins} we show the results for the protein-protein
interaction network of \emph{Escherichia coli}, for which error
estimates in the form of $Q_{ij}$ probabilities are
provided~\cite{szklarczyk_string_2017}. The probabilities are computed
in an elaborate manner by combining seven sources of evidence for the
existence of an interaction between two proteins. As seen in the figure,
our method is able to detect prominent large-scale features that help
shape the posterior distribution. The resulting posterior probabilities
are fairly different from the primary error estimates, showing that
these observed correlations can be very informative for the
reconstruction process.

\section{Conclusion}\label{sec:conclusion}

We have presented a general nonparametric Bayesian network
reconstruction framework that couples a noisy measurement model with the
stochastic block model (SBM) as a generative process. The posterior
distribution of this joint model yields simultaneously an ensemble of
possibilities for the underlying network, as well as its large-scale
hierarchical modular organization. As we have shown, this joint
identification of the network structure enables the existence of
correlations in the measured data to inform the network
reconstruction. As a consequence, our method can be employed also when a
single measurement of the network has been made --- which is not
possible with methods that do not exploit such correlations --- and the
error probabilities are unknown. This property makes our approach
applicable to the dominating set of network datasets that do not
provide primary error estimates of any kind, and can extract from them
not only the most likely underlying network, but also error estimates
for arbitrary network properties.

We have shown that our general methodology is versatile, allowing for
different noise models. We have considered the situation where the error
probabilities are heterogeneous, showing strong evidence for its
existence in empirical data, and demonstrated the efficacy of our
modified approach in capturing it. We have also shown how extraneous
uncertainty estimations obtained with arbitrary methods can be
incorporated into our approach, without requiring a detailed model for
their generation.

The approach we have proposed is open ended, and admits many extensions
and generalizations. For example, although the SBM can be used to
exploit edge correlations if favor of reconstruction, this can be
further improved by considering more realistic models that include other
kinds of correlations such as triadic closure~\cite{strauss_model_1975}
or latent
spaces~\cite{hoff_latent_2002,newman_generalized_2015-1}. Furthermore,
there is a wide range of possibilities for other kinds of noise models
different from the ones considered here, including missing and
duplicated nodes, and edge endpoint swaps (e.g. that can occur from
crossings in imaging data). Additionally, network data often come with a
wealth of node and edge
annotations~\cite{newman_structure_2016,hric_network_2016}, with
important special cases being
weighted~\cite{aicher_learning_2015,peixoto_nonparametric_2018} and
multilayer~\cite{kivela_multilayer_2014,peixoto_inferring_2015}
networks. These extra data are potentially useful for reconstruction,
although they also contain their own measurement errors. Determining the
most appropriate and effective manner to model and exploit this extra
information in reconstruction seems like fertile grounds for future
work.

\begin{acknowledgments}
This research made use of the Balena High Performance Computing (HPC)
Service at the University of Bath.
\end{acknowledgments}

\appendix

\section{Beta prior distribution}\label{app:beta}

\begin{figure}
  \includegraphics{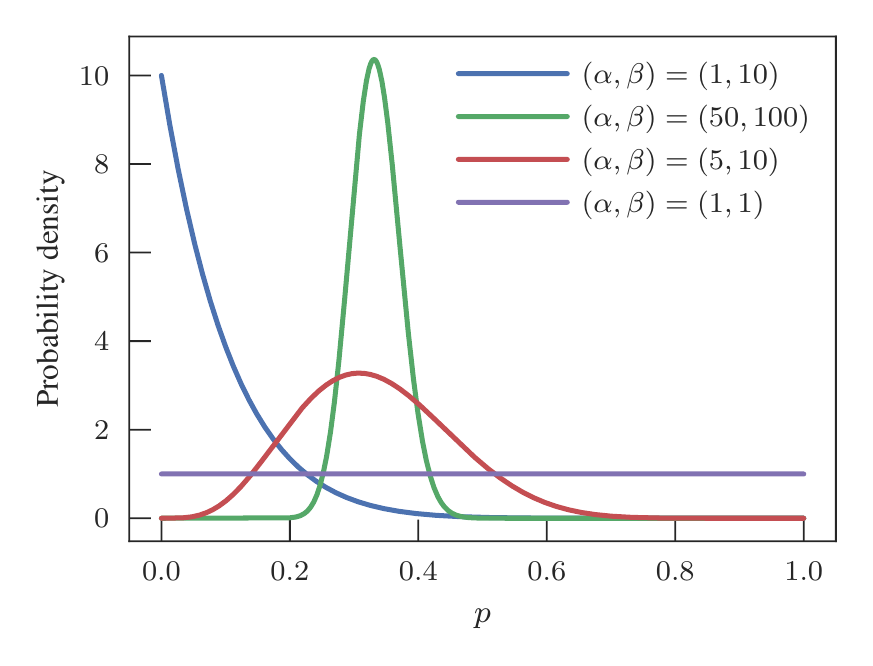}

  \caption{Beta distributions for the noise magnitudes $p$ and $q$ allow
  us to control the degree of prior knowledge we have on their
  values. For example, the values $(\alpha,\beta)=(1,10)$ represent an
  expectation that the value of $p$ is relatively low, with mode at $0$
  and average $\alpha/(\alpha+\beta)=1/11\approx 0.09$. The values
  $(\alpha,\beta)=(50,100)$ express relative certainty that the value of
  $p$ is close to $1/3$, whereas the values $(\alpha,\beta)=(5,10)$
  represent the same average expectation, but with less certainty. The
  values $(\alpha,\beta)=(1,1)$ express the largest amount of
  uncertainty about the parameter $p$, in which case it is uniformly
  distributed in unit interval. \label{fig:beta}}
\end{figure}

In Fig.~\ref{fig:beta} are shown examples of the Beta distribution of
Eq.~\ref{eq:beta}, for different choices of the hyperparameters $\alpha$
and $\beta$, illustrating their meaning with respect to the prior
knowledge assumed for the missing edge probability $p$ (and
analogously for the spurious edge probability $q$, and its
hyperparameters $\mu$ and $\nu$).

\section{Latent edge MCMC algorithm}\label{app:mcmc}

As described in the main text, we use a Markov chain Monte Carlo (MCMC)
algorithm to sample from the posterior distribution
\begin{equation}
  P(\A|\D) = \frac{P(\D|\A)P(\A)}{P(\D)},
\end{equation}
where $\A$ is the network being inferred, and $\D$ is the measurement
data. Since we are using structured distributions in place of $P(\A)$,
consisting of nonparametric formulations of the SBM, its computation in
closed form is not tractable. Instead, we sample from the joint
posterior
\begin{equation}\label{eq:joint_posterior_app}
  P(\A,\bb|\D) = \frac{P(\D|\A)P(\A|\bb)P(\bb)}{P(\D)},
\end{equation}
where $\bb$ is the partition of nodes used for the SBM. If we sample
from this distribution, and ignore the values of $\bb$, we obtain the
desired marginal $P(\A|\D)=\sum_{\bb}P(\A,\bb|\D)$. However, we
are often also interested in the partition itself, as it gives
information on the large-scale network structure, so we often use this
in our analyses as well.

The MCMC algorithm consists of making proposals of the kind
$P(\bb'|\A,\bb)$ and $P(\A'|\A,\bb)$ for the partition and network,
respectively, and accepting them according to the
Metropolis-Hastings probability
\begin{equation}\label{eq:metropolis_app}
  \min\left(1, \frac{P(\A',\bb'|\D)P(\A|\A',\bb')P(\bb|\A',\bb')}{P(\A,\bb|\D)P(\A'|\A,\bb)P(\bb'|\A,\bb)}\right),
\end{equation}
which does not require the computation of the intractable normalization
constant $P(\D)$. In practice, at each step in the chain we make either
a move proposal for $\A$ or $\bb$, not both at once. For the node
partition, we use the move proposals similar to the ones used in
Refs.~\cite{peixoto_efficient_2014,peixoto_nonparametric_2017}, where
for any given node $i$ in group $r$ we propose to move it to group $s$
(which can be previously unoccupied, in which case it is labelled
$s=B+1$) according to
\begin{multline}
  P(b_i=r\to s|\A,\bb) = d\delta_{s,B+1} + {}\\
 (1-d)(1-\delta_{s,B+1})\sum_{t=1}^{B}P(t|i)\frac{e_{ts} + \epsilon}{e_t + \epsilon B},
\end{multline}
where $P(t|i) = \sum_jA_{ij}\delta_{b_j,t}/k_i$ is the fraction of
neighbors of $i$ that belong to group $t$, $\epsilon > 0$ is a small
parameter which guarantees ergodicity, and $d$ is the probability of
moving to a previously unoccupied group. (If $k_i = 0$, we assume
$P(b_i=r\to s|\A,\bb)=d\delta_{s,B+1} + (1-d)(1-\delta_{s,B+1})/B$.)
This move proposal attempts to the use the currently known large-scale
structure of the network to better inform the possible moves of the
node, without biasing with respect to group assortativity. The
parameters $d$ and $\epsilon$ do not affect the correctness of the
algorithm, only the mixing time, which is typically not very sensitive,
provided they are chosen within a reasonable range (we used $d=0.01$ and
$\epsilon=1$ throughout).  When using the HDC-SBM, we used the variation
of the above for hierarchical partitions described in
Ref.~\cite{peixoto_nonparametric_2017}. The move proposals above require
only minimal bookkeeping of the number edges incident on each group, and
can be made in time $O(k_i)$, which is also the time required to compute
the ratio in Eq.~\ref{eq:metropolis_app}, independent on how many groups
are currently occupied.

For the network move proposals we could have used simple edge/nonedge
flips with
\begin{equation}
  P(A_{ij}' = A_{ij} + \delta|\A) =
  \begin{cases}
    1 & \text{ if } A_{ij} + \delta = 1 - A_{ij},\\
    0 & \text{ otherwise},
  \end{cases}
\end{equation}
with $\delta \in \{-1, 1\}$. But in fact, since we operate with latent
multigraphs, the moves are slightly different, as described in
Appendix~\ref{app:multigraphs}. The correctness of the algorithm does not
depend on the order or the frequency with which we attempt to update the
entries $(i,j)$, provided they are all eventually updated, so in
principle we could choose them randomly each time. However, we have
found this leads to poor mixing times, since most entries
correspond to nonedges $A_{ij}=0$ which tend to remain in that
state. Instead, we choose the entries to update with a probability given
by the current SBM,
\begin{equation}\label{eq:move}
  P(i,j|\A,\bb) = \kappa_i\kappa_jm_{b_i,b_j},
\end{equation}
with
\begin{equation}
  \kappa_i = \frac{k_i+1}{\sum_j\delta_{b_j,b_i}k_j+1}
\end{equation}
being the probability of selecting node $i$ from its group $b_j$,
proportional to its current degree plus one, and
\begin{equation}
  m_{rs} = \frac{e_{rs} + 1}{\sum_{tu}e_{rs}+1}
\end{equation}
is the probability of selecting groups $(r,s)$, where
$e_{rs}=\sum_{ij}A_{ij}\delta_{b_i,r}\delta_{b_j,s}$. The above
probabilities guarantee that every entry will be eventually sampled, but
it tends to probe denser regions more frequently, which we found to
typically lead to faster mixing times. This sampling can be done in time
$O(1)$, simply by keeping urns of vertices and edges according to the
group memberships. The time required to compute the ratio in
Eq.~\ref{eq:metropolis_app} is also $O(1)$ for the DC-SBM and $O(L)$ for
the HDC-SBM, where $L$ is the hierarchy depth, again independent of the
number of occupied groups.

When combining both move proposals above for the partition and network,
the time required to perform $V$ node proposals and $M$ edge proposals
is $O(\avg{k}V + M)$, where $\avg{k}$ is the average degree, which
allows for the inference of very large networks, with up to millions of
edges. A reference implementation of the above algorithm is freely
available as part of the \texttt{graph-tool}
library~\cite{peixoto_graph-tool_2014}.

\section{Nonparametric SBM formulation}

Here we give a summary of the nonparametric SBMs used in this work,
which are derived in detail in Ref.~\cite{peixoto_nonparametric_2017}. We
begin with the Poisson DC-SBM likelihood~\cite{karrer_stochastic_2011},
\begin{multline}\label{eq:dc-sbm}
  P(\A|\bm{\lambda},\bm{\theta},\bb) = \prod_{i< j}\frac{e^{-\theta_i\theta_j\lambda_{b_i,b_j}}(\theta_i\theta_j\lambda_{b_i,b_j})^{A_{ij}}}{A_{ij}!}\times\\
  \prod_i\frac{e^{-\theta_i^2\lambda_{b_i,b_i}/2}(\theta_i^2\lambda_{b_i,b_i}/2)^{A_{ii}/2}}{(A_{ii}/2)!},
\end{multline}
which generates multigraphs with $A_{ij}\in\mathbb{N}$, and with
self-loops allowed. By choosing the arbitrary parametrization
$\sum_i\theta_i\delta_{b_i,r}=1$ for every group $r$, $\lambda_{rs}$
becomes the expected number of edges between groups $r$ ans $s$, and
$\theta_i$ is proportional to the expected degree of node $i$,
$\theta_i=\avg{k_i}/\sum_s\lambda_{b_i,s}$. We use the noninformative
prior for $\bm{\theta}$,
\begin{equation}
  P(\bm{\theta}|\bb) = \prod_r(n_r-1)!\delta({\textstyle\sum_i\theta_i\delta_{b_i,r}-1}).
\end{equation}
and $\bm{\lambda}$,
\begin{equation}\label{eq:prior-lambda-dc}
  P(\bm{\lambda}|\bb) = \prod_{r\le s}e^{-\lambda_{rs}/(1+\delta_{rs})\bar{\lambda}}/(1+\delta_{rs})\bar{\lambda}
\end{equation}
with $\bar{\lambda}=2E/B(B+1)$, which results in the integrated marginal
probability,
\begin{align}
  P(\A|\bb) &= \int P(\A|\bm{\lambda},\bm{\theta},\bb)P(\bm{\lambda}|\bb)P(\bm{\theta}|\bb)\;\dd\bm{\lambda}\dd\bm{\theta}\nonumber\\
  &= \frac{\bar\lambda^E}{(\bar{\lambda}+1)^{E+B(B+1)/2}}\times\frac{\prod_{r<s}e_{rs}!\prod_re_{rr}!!}{\prod_{i<j}A_{ij}!\prod_iA_{ii}!!}
  \times \\
  &\qquad\prod_r\frac{(n_r - 1)!}{(e_r + n_r-1)!} \times \prod_ik_i!,\label{eq:dc-sbm-marginal}
\end{align}
where $k_i=\sum_jA_{ij}$ is the degree of node $i$. As shown in
Ref.~\cite{peixoto_nonparametric_2017}, the above is equivalent to a
microcanonical model given by
\begin{equation}\label{eq:dc-ensemble-equivalence}
  P(\A|\bb) = P(\A|\bm{k},\e,\bb)P(\bm{k}|\e,\bb)P(\e|\bb),
\end{equation}
with
\begin{align}
  P(\A|\bm{k},\e,\bb) &= \frac{\prod_{r<s}e_{rs}!\prod_re_{rr}!!\prod_ik_i!}{\prod_{i<j}A_{ij}!\prod_iA_{ii}!!\prod_re_r!!},\label{eq:micro-dc-sbm}\\
  P(\bm{k}|\e,\bb) &= \prod_r\multiset{n_r}{e_r}^{-1},\label{eq:micro-uniform-degrees}\\
  P(\e|\bb) &= \bar\lambda^E/(\bar\lambda+1)^{E+B(B+1)/2},
\end{align}
being the corresponding noninformative priors. Following
Ref.~\cite{peixoto_nonparametric_2017}, we replace the
microcanonical prior for the degrees with
\begin{equation}\label{eq:k_dist}
  P(\bm{k}|\bm{e},\bm{b}) = P(\bm{k}|\bm{\eta})P(\bm{\eta}|\bm{e},\bm{b})
\end{equation}
where $\bm{\eta}=\{\eta_k^r\}$ are the degree frequencies of each group,
with $\eta_k^r$ being the number of nodes with degree $k$ that belong to
group $r$, and
\begin{equation}
  P(\bm{k}|\bm{\eta}) = \prod_r \frac{\prod_k\eta_k^r!}{n_r!}
\end{equation}
is a uniform distribution of degree sequences constrained by the overall
degree counts, and finally
\begin{equation}\label{eq:k_dist_prior}
  P(\bm{\eta}|\bm{e},\bm{b}) = \prod_r q(e_r, n_r)^{-1}
\end{equation}
is the distribution of the overall degree counts. The quantity $q(m,n)$
is the number of different degree counts with the sum of degrees being
exactly $m$ and that have at most $n$ non-zero counts, given by
\begin{equation}\label{eq:q-exact}
  q(m, n) = q(m, n - 1) + q(m - n, n).
\end{equation}
For the node partition we use the prior,
\begin{equation}\label{eq:partition-prior}
  P(\bb) = P(\bb|\bm{n})P(\bm{n}|B)P(B)=\frac{\prod_rn_r!}{N!}{N-1\choose B-1}^{-1}N^{-1}.
\end{equation}
which is agnostic to group sizes.

The hierarchical degree-corrected SBM (HDC-SBM) is obtained by replacing
the uniform prior for $P(\e|\bb)$ by a nested sequence of SBMs, where
the edge counts in level $l$ are generated by a SBM at a level above,
\begin{equation}\label{eq:multi_sbm}
  P(\bm{e}_l|\bm{e}_{l+1},\bm{b}_l) = \prod_{r<s}\multiset{n_r^ln^l_s}{e_{rs}^{l+1}}^{-1}
  \prod_{r}\multiset{n_r^l(n_r^l+1)/2}{e_{rr}^{l+1}/2}^{-1},
\end{equation}
where $\multiset{n}{m} = {n+m-1\choose m}$ is the multiset coefficient.
The prior for the hierarchical partition is obtained using
Eq.~\ref{eq:partition-prior} at every level. We refer to
Ref.~\cite{peixoto_nonparametric_2017} for further details.

Directed variations of the model above are
straightforward~\cite{peixoto_nonparametric_2017}, together with their
noise models considered in the text, which simply require sums and
products to go over all directed node pairs. We omit the expressions
here for brevity, but we used the directed models whenever appropriate.

The hierarchical model above is constructed to be agnostic about several
large-scale aspects of the network, including the degree distribution,
the distribution of group sizes and the mixing patterns. Due to its
nonparametric nature, it can be used to infer the dimensions of the
model, including the number of groups and hierarchy shape. The HDC-SBM
has the additional advantage that it can detect small but statistically
significant groups in large networks, where the maximum number of
detectable groups scales with $O(N/\ln N)$, as opposed to the
$O(\sqrt{N})$ obtainable with non-hierarchical
models~\cite{peixoto_parsimonious_2013,peixoto_hierarchical_2014}.

\section{Adapting multigraph models to simple graphs}\label{app:multigraphs}

The SBM variations considered in the previous section generate
multigraphs with self-loops, however the noise models considered in this
work operate on simple graphs. The usual justification for the use of
multigraph models on simple graph data is that in the sparse case they
are approximately the same, since the probability of multiple edges and
self-loops being generated is very small. Although this is true for
uniform SBMs, like the planted partition model considered in
Sec.~\ref{sec:detectability}, it may not be true for the DC-SBM when the
degree distribution is sufficiently broad. In this situation, the simple
and multigraph ensembles are no longer
equivalent~\cite{park_origin_2003,johnson_entropic_2010,garlaschelli_ensemble_2017},
and the use of the multigraph model in this case may lead to
biases. Unfortunately, the simple graph formulations of the DC-SBM
cannot have their integrated likelihoods computed in closed form.

Here we adapt the multigraph models to simple graphs in tractable and
simple way by generating multigraphs and then collapsing the multiple
edges. In other words, if $\G$ is a multigraph with entries $G_{ij} \in
\mathbb{N}$, the collapsed simple graph $\A(\G)$ has binary entries
\begin{equation}
  A_{ij}(G_{ij}) =
  \begin{cases}
    1 & \text{ if } G_{ij} > 0 \text{ and } i\ne j,\\
    0 & \text{ otherwise.}
  \end{cases}
\end{equation}
Therefore, if $\G$ is a multigraph generated by $P(\G|\theta)$, where
$\theta$ are arbitrary parameters, then the corresponding collapsed simple
graph $\A$ is generated by
\begin{align}
  P(\A | \theta) &= \sum_{\G}P(\A,\G|\theta),\\
  &= \sum_{\G}P(\A|\G)P(\G|\theta),
\end{align}
with
\begin{equation}
  P(\A|\G) =
  \begin{cases}
    1 & \text{ if } \A = \A(\G),\\
    0 & \text{ otherwise.}
  \end{cases}
\end{equation}
Even if $P(\A | \theta)$ cannot be computed in closed form, the joint
distribution $P(\A,\G|\theta) = P(\A|\G)P(\G|\theta)$ is trivial,
provided we have $P(\G|\theta)$ in closed form. Therefore, instead of
directly sampling from the posterior distribution
\begin{equation}
  P(\A,\bb|\D) = \frac{P(\D|\A)P(\A,\bb)}{P(\D)},
\end{equation}
we sample from the joint posterior
\begin{equation}
  P(\A,\G,\bb|\D) = \frac{P(\D|\A)P(\A|\G)P(\G,\bb)}{P(\D)},
\end{equation}
using MCMC, treating the values $G_{ij}$ as latent variables, and then
we marginalize
\begin{equation}
  P(\A,\bb|\D) = \sum_{\G}P(\A,\G,\bb|\D),
\end{equation}
which is done simply by sampling from $P(\A,\G,\bb|\D)$ and ignoring the
actual magnitudes of the $G_{ij}$ values, and the diagonal entries. This
yields an almost identical MCMC algorithm to the one described in
Appendix~\ref{app:mcmc}, with the only difference that we keep track of
the values of $G_{ij}$, which are no longer binary, but automatically
give us $A_{ij}$ [which are used for the computation of $P(\D|\A)$]. The
move proposals of the entries of $G_{ij}$ are done by unity changes,
\begin{equation}
  P(G_{ij}' = G_{ij} + \delta|\G) =
  \begin{cases}
    1/2 & \text{ if } G_{ij} > 0,\\
    1   & \text{ if } G_{ij} = 0 \text{ and } \delta = 1,\\
    0   & \text{ otherwise},
  \end{cases}
\end{equation}
again for $\delta \in \{-1, 1\}$.

In the case of the DC-SBM, the degree correction happens
for the multigraph $\G$, and only indirectly for $\A$. But since our
model is nonparametric, and the degrees of $\G$ are also generated from
their own priors, this gives us a perfectly valid and useful
degree-corrected model for $\A$ as well.

\section{Datasets}\label{app:data}
Here we give brief descriptions of the datasets used in this work, with
properties listed in tables~\ref{tab:empirical} and~\ref{tab:multiple}. 

\subsection{Data without primary error estimates}
\begin{description}
\item[Karate club] Social network between $34$ members of a Karate
club~\cite{zachary_information_1977}. The version used in
Table~\ref{tab:empirical} is the same one used in
Ref.~\cite{girvan_community_2002}, with $A_{23,34}=1$ and hence $78$
edges in total. In Table~\ref{tab:multiple}, it was assumed that each
repeated entry of the adjacency matrix reported in
Ref.~\cite{zachary_information_1977} amounted to a different
measurement, so that $n_{ij}=2$ and $x_{ij}=2A_{ij}$ for all $(i,j)$,
except for $x_{23,34}=1$.
\item[9/11 terrorists] Social associations between 62 terrorists
responsible for the 9/11
attacks~\cite{krebs_mapping_2002,krebs_uncloaking_2002}.
\item[American football] Network of American football games between
Division IA colleges during the regular season in fall of
2000~\cite{girvan_community_2002}.
\item[Network scientists] Coauthorship network of scientists
working on network science~\cite{newman_finding_2006}.
\item[\emph{C. elegans} neural] Directed  neural network of the \emph{Caenorhabditis
elegans} worm~\cite{white_structure_1986}, manually compiled by Watts et
 al.~\cite{watts_collective_1998}, based on the original data. The $5$
 nodes with zero degree omitted in Ref.~\cite{watts_collective_1998}
 were included in our analysis, resulting in $N=302$ nodes in total.
\item[Malaria genes] Bipartite gene-substring association network for
malaria~\cite{larremore_network_2013}.
\item[Power grid] Western states power grid of the United States~\cite{watts_collective_1998}.
\item[Political blogs] Citations between political blogs during the 2004
presidential election in the United States~\cite{adamic_political_2005}.
\item[DBLP citations] Citation network of DBLP, a database of scientific publications~\cite{ley_dblp_2002}.
\item[Openflights] Directed network of flights between world-wide
airports, collected from the community-driven website
\url{http://www.openflights.org}.
\item[Reactome] Network of protein–protein interactions in humans.~\cite{joshi-tope_reactome:_2005}
\item[cond-mat] Network of collaborations in papers published in the
\texttt{cond-mat} section of the \url{arxiv.org} pre-print website in
the period spanning from January 1, 1995 and March 31,
2005~\cite{newman_structure_2001}.
\item[Enron email] Emails sent between employees of Enron between 1999 and 2003~\cite{klimt_introducing_2004}.
\item[Linux source] Network of Linux source code files, with directed
edges denoting that they include each other\cite{kunegis_konect:_2013}.
\item[Brightkite] Online social network from the defunct brightkite website.
\item[PGP] Global web of trust of the Pretty-Good-Privacy (PGP)
encryption protocol. Nodes are public keys, and directed edges indicate
that one key digitally signed another~\cite{richters_trust_2011}.
\item[Internet AS] Directed network of internet autonomous systems,
ca. 2009, as measured by the Center for Applied Internet Data Analysis
(CAIDA), available at \url{https://www.caida.org/data/}.
\item[Web Stanford] Directed network of hyperlinks between the web pages
from the website of the Stanford University~\cite{leskovec_community_2008}.
\item[Flickr] Network of images in the image-sharing site
\url{http://flickr.com}, where two images are connected if they share
metadata, such tags, groups or location~\cite{mcauley_discovering_2014}.
\end{description}

\subsection{Data with primary error estimates}

\begin{description}
\item[Reality mining] Proximity interactions between voluntary students over
time~\cite{eagle_reality_2006}. As measurements we considered the state
of the network during eight consecutive Wednesdays in March and April of
2005.
\item[School friends] Directed network of friendship between primary and
high-school students~\cite{moody_peer_2001}. Each student have been
asked repeatedly to list his or her best 5 female and 5 male friends.
\item[Human connectome] Neuronal connections in the human brain,
  measured for $418$ individuals, each of which we considered
as a separate measurement~\cite{szalkai_parameterizable_2017}.
\end{description}

\bibliography{bib,extra}

\end{document}